%% file: diet-cdf-aistats2023.tex
\documentclass[twoside]{article}

\usepackage[accepted]{aistats2023}

\input{preamble/packages.tex}
\input{preamble/definitions.tex}

\input{preamble/math.tex}

\usepackage{thmtools,thm-restate}
\usepackage{tikzit}
\input{fig/graph.tikzstyles}

\allowdisplaybreaks

\begin{document}

\twocolumn[

\aistatstitle{DIET: Conditional independence testing with  marginal dependence measures of residual information}

\aistatsauthor{ Mukund Sudarshan*\textsuperscript{1} \And Aahlad Puli*\textsuperscript{1} \And  Wesley Tansey\textsuperscript{4} \And Rajesh Ranganath\textsuperscript{1,2,3}\\}

\aistatsaddress{\textsuperscript{1}Computer Science, \textsuperscript{2}Data Science, and \textsuperscript{3}Population Health at Langone Health, New York University
\\ \textsuperscript{4}Computational Oncology Memorial Sloan Kettering Cancer Center} ]

\begin{abstract}
\input{sections/abstract.tex}

\end{abstract}

\section{INTRODUCTION}
\input{sections/intro.tex}

\section{BACKGROUND}
\label{sec:background}

\input{sections/background.tex}

\section{DIET}
\input{sections/diet-cdf.tex}

\section{THEORETICAL ANALYSIS OF DIET}

\input{sections/diet-theoretical-properties.tex}

\section{EXPERIMENTS}
\input{sections/experiments.tex}

\section{DISCUSSION}

\input{sections/discussion.tex}

\section*{Acknowledgements}

We thank the reviewers for their thoughtful comments.
We would like to thank the participants of the Selective Inference Seminar for their helpful feedback: in particular Lihua Lei, Rina Barber, and Lucas Janson.
This work was supported by the PhRMA Foundation Predoctoral Fellowship, NIH/NHLBI Award R01HL148248, NSF Award 1922658 NRT-HDR: FUTURE Foundations, Translation, and Responsibility for Data Science, NSF CAREER Award 2145542, NIH U54CA274492, R37CA271186, Break Through Cancer, and the Tow Center for Developmental Oncology.

\bibliography{citations.bib}
\bibliographystyle{abbrvnat}

\onecolumn

\appendix

\section{APPENDIX}
\input{sections/appendix_proofs.tex}

\section{EXPERIMENTAL DETAILS}
\input{sections/appendix_experimental_details.tex}

\end{document}

%% file: preamble/packages.tex
\usepackage{amssymb, amsmath, amsthm}
\usepackage{nccmath}

\usepackage[english]{babel}
\usepackage{enumitem}
\usepackage{wrapfig}
\usepackage{multicol}
\usepackage{float}
\usepackage[utf8]{inputenc} %
\usepackage[T1]{fontenc}
\usepackage{amsfonts}
\usepackage{pifont}
\usepackage{bm}
\usepackage{bbm}
\usepackage{graphicx}
\usepackage[font=small,labelfont=bf]{caption}
\usepackage[format=hang]{subcaption}
\usepackage{booktabs}
\usepackage[ruled,vlined]{algorithm2e}
\SetKwInput{KwInput}{Input}
\SetKwInput{KwOutput}{Output}
\usepackage{listings}
\usepackage{fancyvrb}
\fvset{fontsize=\normalsize}
\usepackage[round]{natbib}
\usepackage[acronym,smallcaps,nowarn]{glossaries}
\glsdisablehyper
\usepackage[colorlinks=true,linktoc=all]{hyperref}
\usepackage[all]{hypcap}
\usepackage{xcolor}
\definecolor{deepblue}{HTML}{20639B}
\definecolor{seagreen}{HTML}{3CAEA3}
\definecolor{niceyellow}{HTML}{F6D55C}

\hypersetup{citecolor=deepblue}
\hypersetup{linkcolor=deepblue}
\usepackage[nameinlink]{cleveref}

\usepackage{multirow}

\usepackage{nicefrac}

%% file: preamble/definitions.tex
\newacronym{crt}{crt}{conditional randomization test}
\newacronym{dcrt}{dcrt}{distilled conditional randomization test}
\newacronym{snp}{snp}{single nucleotide polymorphism}
\newacronym{fdr}{fdr}{false discovery rate}
\newacronym{cpt}{cpt}{conditional permutation test}
\newacronym{fwer}{fwer}{family-wise error rate}
\newacronym{fpr}{fpr}{false positive rate}
\newacronym{fdp}{fdp}{false discovery proportion}
\newacronym{tpr}{tpr}{true positive rate}
\newacronym{mle}{mle}{maximum likelihood estimation}
\newacronym{kld}{kl}{Kullback–Leibler divergence}
\newacronym{pt}{pt}{permutation test}
\newacronym{shap}{shap}{shapley additive explanations}
\newacronym{roc}{roc}{receiver operating characteristic}
\newacronym{fcauc}{fcauc}{\textsc{fdr}-controlled area under the \textsc{roc} curve}
\newacronym{rbf}{rbf}{radial basis function}
\newacronym{rf}{rf}{random forest}
\newacronym{ddlk}{ddlk}{deep direct likelihood knockoffs}
\newacronym{hrt}{hrt}{holdout randomization test}
\newacronym{ourmodel}{ddlk}{deep direct likelihood knockoffs}
\newacronym{gan}{gan}{generative adversarial network}
\newacronym{mmd}{mmd}{maximum mean discrepancy}
\newacronym{mdn}{mdn}{mixture density network}
\newacronym{cbc}{cbc}{complete blood count}
\newacronym{vae}{vae}{variational auto-encoder}
\newacronym{hmm}{hmm}{hidden markov model}
\newacronym{mcmc}{mcmc}{markov chain monte carlo}
\newacronym{cvs}{cvs}{controlled variable selection}
\newacronym{ci}{ci}{conditional independence}
\newacronym{ourmethod}{contra}{contrarian randomization test}
\newacronym{bh}{bh}{Benjamini-Hochberg}
\newacronym{cdf}{cdf}{cumulative distribution function}
\newacronym{pdf}{pdf}{probability density function}
\newacronym{mvn}{mvn}{multivariate normal}
\newacronym{diet}{diet}{decoupled independence test}
\newacronym{diet-cdf}{diet-cdf}{USE ONLY WITH ACRSHORT}
\newacronym{gcit}{gcit}{generative conditional independence test}
\newacronym{gcm}{gcm}{generalized covariance measure}
\newacronym{iv}{iv}{instrumental variable}
\newacronym{ehr}{ehr}{electronic health record}

\newtheorem{theorem}{Theorem} %
\newtheorem{corollary}{Corollary}[theorem]

\crefname{thmdef}{def.}{defs.}
\Crefname{thmdef}{Definition}{Definitions}
\crefname{lemma}{lemma}{lemmas}
\crefname{prop}{prop.}{props.}
\crefname{algorithm}{algorithm}{algorithms}
\Crefname{algorithm}{Algorithm}{Algorithms}
\crefname{algocf}{Alg.}{Algs.}
\Crefname{algocf}{Algorithm}{Algorithms}
\crefname{appsec}{app.}{apps.}
\Crefname{appsec}{A.}{A.}
\newif\ifprintcomments

%% file: preamble/math.tex
\def\eqref#1{equation~\ref{#1}}

\def\1{\bm{1}}

\def\rva{{\mathbf{a}}}
\def\rvb{{\mathbf{b}}}

\def\rvd{{\mathbf{d}}}
\def\rve{{\mathbf{e}}}

\def\rvp{{\mathbf{p}}}

\def\rvx{{\mathbf{x}}}
\def\rvxt{{\widetilde{\mathbf{x}}}}
\def\rvy{{\mathbf{y}}}
\def\rvz{{\mathbf{z}}}
\def\rvzt{{\widetilde{\mathbf{z}}}}

\def\rmXt{{\widetilde{\mathbf{X}}}}

\DeclareMathAlphabet{\mathsfit}{\encodingdefault}{\sfdefault}{m}{sl}
\SetMathAlphabet{\mathsfit}{bold}{\encodingdefault}{\sfdefault}{bx}{n}

\def\gC{{\mathcal{C}}}
\def\gD{{\mathcal{D}}}

\def\gH{{\mathcal{H}}}

\def\gO{{\mathcal{O}}}

\def\sP{{\mathbb{P}}}

\def\sR{{\mathbb{R}}}

\newcommand{\E}{\mathbb{E}}

\newcommand{\CI}{\mathrel{\text{{$\perp\hspace{-6pt}\perp$}}}}

\DeclareMathOperator*{\argmax}{arg\,max}
\DeclareMathOperator*{\argmin}{arg\,min}

\newcommand{\g}{\mid}
\newcommand{\D}{\mathcal{D}}

\newcommand{\balpha}[0]{\boldsymbol{\alpha}}

\newcommand{\balphai}[0]{\boldsymbol{\alpha}^{(i)}}
\newcommand{\balphaim}[1]{\boldsymbol{\alpha}^{(i, #1)}}

\newcommand{\qknockoff}{\hat{q}_{\text{knockoff}}}

\newcommand{\qmodel}{\hat{q}_{\text{model}}}

\newcommand{\rvxi}{\rvx^{(i)}}
\newcommand{\rvyi}{\rvy^{(i)}}
\newcommand{\rvzi}{\rvz^{(i)}}
\newcommand{\rvxti}{\rvxt^{(i)}}

\newcommand{\pval}[0]{\hat{\rvp}}

\newcommand{\bdelta}[0]{\bm{\delta}}
\newcommand{\beps}[0]{\bm{\epsilon}}
\newcommand{\dhat}[0]{\smash{\hat{\bm{\delta}}}}
\newcommand{\ehat}[0]{\smash{\hat{\bm{\epsilon}}}}
\newcommand{\dhati}[0]{\hat{\bm{\delta}}^{(i)}}
\newcommand{\ehati}[0]{\hat{\bm{\epsilon}}^{(i)}}

\newcommand{\disteq}[0]{\overset{d}{=}}

\newcommand{\qhat}[0]{\hat{q}}

\newcommand{\qcdf}[0]{\smash{\hat{Q}_{\textsc{cdf}}}}
\newcommand{\qpdf}[0]{\hat{q}_{\textsc{pdf}}}

\newcommand{\Qcdf}[1]{\smash{\hat{Q}_{\textsc{cdf}(#1)}}}

\newcommand{\fzi}[0]{\bar{f}_{\rvzi}}
\newcommand{\gzi}[0]{\bar{g}_{\rvzi}}
\newcommand{\fz}[0]{\bar{f}_{\rvz}}
\newcommand{\gz}[0]{\bar{g}_{\rvz}}

\newcommand{\Dxyz}[0]{\smash{\gD_{\rvx, \rvy, \rvz}}}
\newcommand{\Dxtyz}[0]{\smash{\gD_{\rvxt, \rvy, \rvz}}}

\newcommand{\Dde}[0]{\gD_{\ehat, \dhat}}
\newcommand{\Dbde}[0]{\gD_{\beps, \bdelta}}

\newcommand{\ZT}{M_1}
\newcommand{\ZY}{M_y}

%% file: fig/graph.tikzstyles
\tikzstyle{red node}=[fill=red, tikzit category=nodes, shape=circle, draw=black]
\tikzstyle{white node}=[fill={rgb,255: red,255; green,255; blue,255}, tikzit category=nodes, shape=circle, draw=black]
\tikzstyle{gray node}=[fill={rgb,255: red,221; green,221; blue,221}, tikzit category=nodes, shape=circle, draw=black]
\tikzstyle{blue node}=[fill=blue, shape=circle, draw=black, tikzit category=nodes]
\tikzstyle{green node}=[tikzit fill=green, fill=green, shape=circle, draw=black, tikzit category=nodes]
\tikzstyle{yellow square}=[draw=black, fill=yellow, shape=rectangle]
\tikzstyle{blue node 2}=[fill={rgb,255: red,128; green,0; blue,128}, draw=black, shape=circle, tikzit fill=blue]
\tikzstyle{white square}=[fill=white, draw=black, shape=rectangle]
\tikzstyle{white square no border}=[draw=black, shape=rectangle]

\tikzstyle{dashed edge}=[<->, dashed]
\tikzstyle{blue pointer}=[->, draw=blue]
\tikzstyle{black pointer}=[draw={rgb,255: red,0; green,0; blue,0}, ->]
\tikzstyle{thicc black pointer}=[draw={rgb,255: red,0; green,0; blue,0}, ->, line width=0.3mm]
\tikzstyle{undirected thicc black}=[-, draw={rgb,255: red,0; green,0; blue,0}, line width=0.3mm]
\tikzstyle{undirected thicc black dashed}=[-, dashed, draw={rgb,255: red,0; green,0; blue,0}, line width=0.3mm]

%% file: sections/abstract.tex
\Glspl{crt} assess whether a variable $\rvx$ is predictive of another variable $\rvy$, having observed covariates $\rvz$.
\Glspl{crt} require fitting a large number of predictive models, which is often computationally intractable.
Existing solutions to reduce the cost of \glspl{crt} typically split the dataset into a train and test portion, or rely on heuristics for interactions, both of which lead to a loss in power.
We propose the \gls{diet}, an algorithm that avoids both of these issues by leveraging marginal independence statistics to test conditional independence relationships.
\Gls{diet} tests the marginal independence of two random variables: $F_{\rvx \mid \rvz}(\rvx \mid \rvz)$ and $F_{\rvy \mid \rvz}(\rvy \mid \rvz)$ where $F_{\cdot\mid \rvz}(\cdot \mid \rvz)$ is a conditional \gls{cdf} for the distribution $p(\cdot\mid \rvz)$.
These variables are termed ``information residuals.''
We give sufficient conditions for \gls{diet} to achieve finite sample type-1 error control and power greater than the type-1 error rate.
We then prove that when using the mutual information between the information residuals as a test statistic, \gls{diet} yields the most powerful conditionally valid test.
Finally, we show \gls{diet} achieves higher power than other tractable \glspl{crt} on several synthetic and real benchmarks.

%% file: sections/intro.tex
A key question in many scientific disciplines is whether a variable $\rvx$ causes some outcome $\rvy$ \citep{lauritzen1996graphical,pearl2009causal}.
In genetics for example, scientists test whether a particular gene causes cancer to design targeted therapies \citep{zhu2018causal}. 
When there are confounders $\rvz$ that may affect both $\rvx$ and $\rvy$, assessing the causal link between $\rvx$ and $\rvy$ corresponds to testing the \gls{ci} between $\rvx$ and $\rvy$ given $\rvz$:
\begin{align}
\begin{split}
	\text{null hypothesis } &\mathcal{H}_0 : \rvx \CI \rvy \mid \rvz
	\\
	\text{alternate hypothesis } &
	\mathcal{H}_1 : \rvx \not \CI \rvy \mid \rvz .
	\label{eqn:ci-hypothesis}
\end{split}
\end{align}
The advantage of using a hypothesis test for understanding such relationships is the ability to explicitly control the type-1 error rate: the probability of erroneously 
rejecting the null hypothesis where $\rvx$ is independent of $\rvy$ conditioned on $\rvz$.
Consequently, constructing conditional independence hypothesis tests has become increasingly popular in the machine learning literature \citep{zhang2012kernel,doran2014permutation,sen2017model,runge2018conditional,bellot2019conditional}.

Many existing tests however, have been shown to lose power when the dimensionality of $\rvz$ is high due to reliance on kernels \citep{bellot2019conditional} or fail to control the type-1 error rate when strong parametric assumptions about $p(\rvy \mid \rvx, \rvz)$ are violated \citep{candes2018panning}.

To test for conditional independence when $\rvz$ is high-dimensional and without making assumptions on the form of $p(\rvy \mid \rvx, \rvz)$, \citet{candes2018panning} proposed the \acrfull{crt}.
The \gls{crt} calculates a $p$-value for \cref{eqn:ci-hypothesis} by repeatedly comparing a scalar-valued test statistic $T(\Dxyz)$ with draws from the null distribution $T(\Dxtyz^{(m)})$:
\begin{align}
	\frac{1}{M + 1}
	\left(
		1 + \sum_{m=1}^M \mathbbm{1}(T(\Dxyz) \leq T(\Dxtyz^{(m)}))
	\right),
	\label{eqn:crt-pvalue}
\end{align}
where $\Dxyz$ is a set of $N$ iid samples drawn from $p(\rvx$, $\rvy, \rvz)$.
Null samples $\Dxtyz^{(m)}$ are drawn from the distribution $p(\rvz , \rvy) p(\rvx \mid \rvz)$, where $\rvxt \sim p(\rvx \mid \rvz)$ is by construction conditionally independent of $\rvy$ given $\rvz$.
If the null hypothesis is true, then $T(\Dxyz)$ will have the same distribution as each $T(\Dxtyz^{(m)})$.

In contrast with other conditional independence testing methods, the \gls{crt} assumes the ability to sample $p(\rvx \mid \rvz)$ but makes no assumptions on the form of $p(\rvy \mid \rvx, \rvz)$ or the test statistic $T$ to control the type-1 error.
This flexibility enables the use of powerful predictive models and empirical risk test statistics \citep{tansey2018holdout,liu2020fast,sudarshan2021contra} that lead to higher power and better type-1 error rates than classical methods.

However, \glspl{crt} are computationally expensive.
For each null sample, the test statistic must be recomputed.
When using predictive models in empirical risk test statistics, these models must correspondingly be refit for \textit{every} null sample $\Dxtyz^{(m)}$.
When the predictive models are computationally expensive to train, such as deep neural networks, the burden of running a \gls{crt} can become prohibitive.

\paragraph{Related work.}

There are two classes of conditional independence testing methods. 
These can be characterized by the assumptions they make to guarantee type-1 error control.
We term the first class ``Model-Y'' methods, which make assumptions about the $\rvy \mid \rvx, \rvz$ distribution to ensure type-1 error control.
This includes procedures that test for edges in Bayesian networks \citep{koller2009probabilistic, spirtes2000causation, cheng1998learning, de2000new}, kernel-based methods \citep{fukumizu2007kernel, zhang2012kernel}, permutation-based methods \citep{gretton2012kernel, doran2014permutation, lee2017kernel}, and many others.

The other class of conditional independence tests are ``Model-X'' methods, introduced by \citet{candes2018panning}.
These require no assumptions about $\rvy \mid \rvx, \rvz$, but assume access to samples from $\rvx \mid \rvz$.
This approach is more effective in controlling type-1 error than Model-Y methods when the number of labeled samples $(\rvy, \rvx, \rvz)$ is small, but a large unlabeled dataset of $(\rvx, \rvz)$ is available: like in the case of genetics.
Much of the existing work on Model-X methods focuses on how to model $p(\rvx \mid \rvz)$ from data \citep{romano2020deep,sudarshan2020deep,jordon2018knockoffgan}, but leaves to the practitioner the form of the \gls{crt} test statistic $T(\Dxyz)$.

As a result, recent work in the Model-X space focuses on creating powerful but tractable \gls{crt} test statistics.
\citet{liu2020fast} propose a pair of methods called \glspl{dcrt}.
The first method, the $d_0$-\gls{crt} constructs a \gls{crt} where the test statistic is the marginal dependence between $(\rvy - \E[ \rvy \mid \rvz])$ and $(\rvx - \E[\rvx \mid \rvz])$.
However, \citet{liu2020fast} demonstrate empirically that the $d_0$-\gls{crt} achieves low power when $\rvy$ is a function of some non-linear interaction between $\rvx$ and $\rvz$.
To account for this issue, the authors also introduce the $d_I$-\gls{crt}.
The $d_I$-\gls{crt} first uses a heuristic to select a small subset of $\rvz$ to explicitly construct a set of interaction terms with $\rvx$.
It then fits a model $\qhat_{d_I}$ to estimate the conditional expectation of $\rvy$ given $(\rvx - \E[\rvx \mid \rvz]), \E[\rvy \mid \rvz]$, and each of the interaction terms.
The $d_I$-\gls{crt} test statistic is some measure of feature importance of $\rvx - \E[\rvx \mid \rvz]$ in $\qhat_{d_I}$.
If the heuristic pre-selection step fails to select the interactions that occur in the data, the $d_I$-\gls{crt} can fail to achieve power due to its reliance on conditional expectations.

The \gls{hrt} \citep{tansey2018holdout} is another tractable yet flexible \gls{crt}.
It splits samples of data into train and test sets, fits a predictive model on the train set, then uses this model to run a \gls{crt} only on the test set. 
While the \gls{hrt} does not require heuristics for interactions between $\rvx$ and $\rvz$, it often loses power compared to \glspl{dcrt} in practice due to sample splitting between the training and test set \citep{liu2020fast}.

\citet{patra2016nonparametric} develop the notion of a nonparametric residual and study its use in testing for conditional independence, but do not provide a method with guarantees of power or type-1 error control.
Residuals, like the ones computed in $d_0$-\gls{crt} and \gls{diet}, play a part in causal effect estimation under unobserved confounding.
Objects called control functions are estimated as residuals from data and used to adjust for confounding: \citet{guo2016control} use additive residuals, \citet{imbens2009identification} use conditional \glspl{cdf} like in \gls{diet}, and \citet{puli2020general} give a general recipe to construct control functions with identification guarantees.

\paragraph{Our contributions.}

We propose a novel \gls{crt} to test $\rvx \CI \rvy \mid \rvz$ that achieves high power without sample splitting.
\Gls{diet} first estimates two conditional \glspl{cdf}: $F_{\rvx \mid \rvz}(\cdot \mid \cdot)$ and $F_{\rvy \mid \rvz}(\cdot \mid \cdot)$ using a dataset of samples from $p(\rvx, \rvy, \rvz)$.
It then tests $\rvx \CI \rvy \mid \rvz$ by testing the marginal independence of the univariate random variables produced by applying the conditional \glspl{cdf} to $(\rvx, \rvz)$ and $(\rvy, \rvz)$ respectively: $F_{\rvx \mid \rvz}(\rvx \mid \rvz)$ and $F_{\rvy \mid \rvz}(\rvy \mid \rvz)$.
\gls{diet} is computationally simple and, as we show, provides the ability to control type-I error regardless of the data generating distribution $p(\rvx, \rvy, \rvz)$.
Further, we characterize distributions for which \gls{diet} can provably achieve power to correctly reject the null hypothesis.

Then, we discuss the limits of distillation procedures like \gls{diet} or the \glspl{dcrt}: we highlight challenges a general procedure that distills a \gls{ci}-test into a marginal one faces while maintaining type-I error control.
After proving that further assumptions are necessary to overcome the challenges, we characterize conditions that allow one to reason about when a general distillation procedure provably achieves power.
Finally, we validate \gls{diet} empirically on synthetic and real benchmarks and observe that it achieves higher power than several baselines while still controlling the type-1 error rate.

%% file: sections/background.tex
\paragraph{\Acrfullpl{crt}.}

\Glspl{crt} outline a general procedure to test for the conditional independence of two variables $\rvx, \rvy \in \sR$ given covariates $\rvz \in \sR^p$.
Using a dataset of $N$ samples $\Dxyz \in (\sR \times \sR \times \sR^p)^N$ and a function $T: (\sR \times \sR \times \sR^p)^N \to \sR$, they compute the \textit{test statistic} $T(\Dxyz)$.
They then create \textit{null datasets} $\Dxtyz^{(m)} \in (\sR \times \sR \times \sR^p)^N$ by copying $\Dxyz$ and replacing $\rvx$ with new samples of $\rvxt \sim p(\rvx \mid \rvz)$\footnote{All \glspl{crt} assume the ability to sample from $p(\rvx \mid \rvz)$ to control type-1 error rates.} to compute \textit{null statistics} $T(\Dxtyz^{(1)}), \dots, T(\Dxtyz^{(M)})$.
Finally, \glspl{crt} use the true and null statistics to compute the $p$-value in \cref{eqn:crt-pvalue}.

\Glspl{crt} in the most general case compute $T(\Dxyz)$ by fitting and then evaluating the performance of a model $\qmodel(\rvy \mid \rvx, \rvz)$ in predicting $\rvy \mid \rvx, \rvz$ \citep{tansey2018holdout,liu2020fast}.
To compute each null statistic, another model is fit and evaluated on each null dataset.
$M+1$ separate models must be fit because \glspl{crt} require that the same function $T$ must be applied to both the true data and the null data.
Given a user-specified \gls{fdr} $\alpha$ and $d$ \gls{ci} tests, $M$ is chosen to be $\gO(\frac{d}{\alpha})$.
For example, at a standard choice of $\alpha = 0.05$, and with just $100$ variables, at least 2000 models need to be fit.
This makes \glspl{crt} intractable. 

In the next section, we introduce \gls{diet}: a flexible \gls{crt} that avoids sample splitting and heuristics like pre-selecting interaction terms.

%% file: sections/diet-cdf.tex
Here we introduce a novel approach to distillation to create a tractable and powerful \gls{crt}.
This section details the construction of the test statistic $T(\Dxyz)$, which measures the marginal dependence between $F_{\rvx \mid \rvz}(\rvx \mid \rvz)$ and $F_{\rvy \mid \rvz}(\rvy \mid \rvz)$.
It then details the computation of each 
 null statistic  $T(\Dxtyz^{(m)})$.
Using the test and null statistics, \gls{diet} computes a $p$-value for testing $\rvx \CI \rvy \mid \rvz$.

\paragraph{Fitting conditional \gls{cdf} estimators.}
Let the \glspl{cdf} associated with the distributions $p(\rvx \mid \rvz)$ and $p(\rvy \mid \rvz)$ be $F_{\rvx \mid \rvz}(\cdot \mid \cdot)$ and $F_{\rvy \mid \rvz}(\cdot \mid \cdot)$ respectively.
\Gls{diet}
tests the marginal independence of the univariate random variables produced by applying the conditional \glspl{cdf} to $(\rvx, \rvz)$ and $(\rvy, \rvz)$ respectively: $F_{\rvx \mid \rvz}(\rvx \mid \rvz)$ and $F_{\rvy \mid \rvz}(\rvy \mid \rvz)$.
As a first step, \gls{diet} estimates these conditional \glspl{cdf} with two estimators: $\Qcdf{{\rvx \mid \rvz}}(\cdot \mid \cdot \,\, ; \theta)$ and $\Qcdf{{\rvy \mid \rvz}}(\cdot \mid \cdot \,\,; \eta)$.
Any conditional \gls{cdf} estimation technique can be used.
Flexible examples include kernel-based methods \citep{bhattacharya1990kernel}, nonparametric estimators, \citep{li2008nonparametric}, and \glspl{mdn} \citep{bishop1994mixture}.
We describe \gls{diet} with \glspl{mdn}.

An \gls{mdn} learns a neural network function $g: \rvz \mapsto \{ \pi_\eta(\rvz)[k], \mu_\eta(\rvz)[k], \sigma_\eta(\rvz)[k] \}_{k=1}^K$ to map values of $\rvz$ to the parameters of a gaussian mixture with $K$ mixture components:
\begin{align*}
	\Qcdf{{\rvy \mid \rvz}}(\rvy \mid \rvz; \eta) = \sum_{k=1}^K \pi_\eta(\rvz)[k] \Phi \left(\frac{\rvy  - \mu_\eta(\rvz)[k]}{ \sigma_\eta(\rvz)[k] } \right ).
\end{align*}
The parameters $\eta$ of $\Qcdf{{\rvy \mid \rvz}}(\rvy \mid \rvz; \eta)$ are learned via maximum likelihood estimation by optimizing over $(\rvy, \rvz)$ pairs in dataset $\Dxyz$:
\begin{align}
	\argmax_{\eta} \frac{1}{N} \sum_{i=1}^N \log \qpdf(\rvy = \rvyi \mid \rvz = \rvzi ; \eta), \label{eqn:dhat-mle}
\end{align}
where $\qpdf$ is the conditional density implied by $\qcdf$.
\Glspl{mdn} are useful as both the conditional \gls{cdf} and density can be computed easily.

A model for $F_{\rvx \mid \rvz}(\cdot \mid \cdot)$, $\Qcdf{{\rvx \mid \rvz}}(\cdot \mid \cdot \,\,  ; \theta)$, is fit similarly but instead of using pairs of $(\rvx, \rvz)$ from $\Dxyz$, \gls{diet} uses only $\rvz$ from $\Dxyz$ and draw samples of $\rvxt \sim p(\rvx \mid \rvz)$ for each $\rvz$ data point.
Note that 
the distribution of $(\rvxt, \rvz)$ is equal to that of $(\rvx, \rvz)$, so evaluating $\Qcdf{{\rvx \mid \rvz}}(\cdot \mid \cdot \,\, ; \theta)$ on samples of $(\rvx, \rvz)$ from $\Dxyz$ will still be in-distribution.

\paragraph{Computing the test statistic $T(\Dxyz)$.}

The \gls{diet} test statistic measures the marginal dependence between two quantities $\ehat$ and $\dhat$ using a dataset of paired samples $\Dde$.
The variables $\ehat$ and $\dhat$, termed ``information residuals'' represent the residual information contained in $\rvx \mid \rvz$ and $\rvy \mid \rvz$.
They are computed as follows.
A sample of $\ehat$ is generated by evaluating the conditional \gls{cdf} $\Qcdf{{\rvx \mid \rvz}}(\cdot \mid \cdot\,\, ; \theta)$ at a sample $(\rvx, \rvz)$, i.e. $\ehat \gets \Qcdf{{\rvx \mid \rvz}}(\rvx \mid \rvz\,\, ; \theta)$.
Similarly, $\dhat \gets \Qcdf{{\rvy \mid \rvz}}(\rvy \mid \rvz ; \eta)$.
To generate the dataset $\Dde$, a pair of $(\ehat, \dhat)$ samples are computed for each $(\rvx, \rvy, \rvz)$ sample in $\Dxyz$ using the respective conditional \glspl{cdf}.

Using the dataset of information residuals $\Dde$, \gls{diet} measures the marginal dependence between $\ehat$ and $\dhat$ using
the estimator of mutual information from \citet{vinh2009information}.
In practice, any measure of dependence $\rho: (\sR \times \sR)^N \to \sR$ can be used.

\input{sections/diet_algorithm.tex}

\paragraph{Computing null statistics $T(\Dxtyz^{(m)})$.}

Computing each null statistic is very similar to computing the test statistic. 
First, a null dataset $\Dxtyz^{(m)}$ is sampled by copying $\Dxyz$, then replacing the $\rvx$ values with $\rvxt \sim p(\rvx \mid \rvz)$.
The same $\qcdf$ models are used to generate information residuals using the null data, after which their mutual information is estimated.
This process is repeated $M$ times to generate $M$ null statistics.

\paragraph{Computing a $p$-value.}

Using the test statistic $T(\Dxyz)$ and each null statistic $T(\Dxtyz^{(m)})$, \gls{diet} computes a $p$-value using \cref{eqn:crt-pvalue}.
The full algorithm is summarized in \cref{alg:diet}.

While the \gls{diet} algorithm is relatively straightforward, it is not obvious why \gls{diet} should control the type-1 error rate, or achieve power.
In the next section, we explore the theoretical properties of \gls{diet}.

%% file: sections/diet_algorithm.tex
\begin{algorithm}[t]
   \caption{\Acrfull{diet}}
   \label{alg:diet}
\SetAlgoLined
\DontPrintSemicolon
\KwInput{Labeled dataset $\Dxyz$,
marginal dependence statistic $\rho$}
\KwOutput{$p$-value $\pval$}
Generate null dataset $\Dxtyz$ by replacing each $\rvx$ in $\Dxyz$ with a sample from $p(\rvx \mid \rvz)$ \;
Fit $\Qcdf{{\rvx \mid \rvz}}(\rvx \mid \rvz ; \theta)$ and $\Qcdf{{\rvy \mid \rvz}}(\rvy \mid \rvz ; \eta)$ using $(\rvx, \rvz)$ pairs and $(\rvy, \rvz)$ pairs from $\Dxtyz$ \;
Generate null datasets $\{ \Dxtyz^{(m)} \}_{m=1}^M$ \;

Create information residual dataset $\Dde$ by evaluating both $\qcdf$ models on $\Dxyz$ \;
\For{$m \in \{1, \dots, M\}$}{
	Create null information residual dataset $\Dde^{(m)}$ by evaluating both $\qcdf$ models on $\Dxtyz^{(m)}$ \;
}
$\pval \gets \frac{1}{M + 1} \left( 1 + \sum_{m=1}^M \mathbbm{1}\left[ \rho(\Dde) \geq \rho(\Dde^{(m)}) \right] \right)$ \;
\end{algorithm}

%% file: sections/diet-theoretical-properties.tex
Here we show that \gls{diet} achieves type-1 error control regardless of the data distribution.
We then discuss when \gls{diet} can provably achieve power and characterize distributions where \gls{diet} is the most powerful test one can perform.
The final part of this section provides a more general perspective on when distillation of a conditional randomization test into a marginal one is possible.
We discuss how assumptions on the data generating process are always needed to guarantee power in a distillation procedure.

\subsection{When can \gls{diet} control the type-1 error rate?}
\label{sec:diet-type-1-error}

The type-1 error rate is the probability that the null hypothesis $\gH_0$ is erroneously rejected: i.e. it is rejected when in reality $\rvx \CI \rvy \mid \rvz$.
To control this error rate at a user-specified level, the $p$-value under $\gH_0$ must either be distributed uniformly over $[0, 1]$ or stochastically dominate\footnote{A random variable $\rva$ stochastically dominates a random variable $\rvb$ if the following partial ordering exists on the \glspl{cdf} of $\rva$ and $\rvb$: $\forall x: F_\rva(x) \leq F_{\rvb}(x)$.} a $\textrm{Uniform}(0,1)$ random variable (see appendix A.1 of \citet{sudarshan2021contra} for a proof of this fact).
\Cref{prop:type-1-error} shows that \acrshort{diet} $p$-values computed using \cref{alg:diet} will stochastically dominate a $\textrm{Uniform}(0,1)$ random variable.

\begin{restatable}{prop}{dieterrorprop}
    Let $(\rvx, \rvy, \rvz)$ be drawn from any distribution $p(\rvx, \rvy, \rvz)$ and $\Dxyz$
    consist of $N$ iid samples from this distribution.
    If $\rvx \CI \rvy \mid \rvz$, then for any measure of marginal dependence $\rho: (\sR \times \sR)^N \to \sR$ the \gls{diet} $p$-value computed using \cref{alg:diet} 
    will stochastically dominate a $\textrm{Uniform}(0,1)$ random variable.
    \label{prop:type-1-error}
\end{restatable}

We detail the full proof in \cref{sec:proof-type-1-error}, but provide a sketch here.
Under $\gH_0$, the test statistic $T(\Dxyz)$ is exchangeable with each of the null statistics $T(\Dxtyz^{(m)})$.
As a result, the $p$-value $\pval$ computed using \cref{eqn:crt-pvalue} will be
uniformly distributed over the set $\{\frac{1}{M + 1}, \frac{2}{M + 1}, \dots, 1\}$.
Such a $p$-value stochastically dominates a Uniform(0, 1) random variable.
\Cref{prop:type-1-error} ensures that if the practitioner rejects the null hypothesis when $\pval \leq \alpha$, the probability of an erroneous rejection is no greater than the significance level $\alpha$.

\subsection{When can \gls{diet} provably achieve power?}

A \gls{crt} achieves power when the distribution of $T(\Dxyz)$ is distinguishable from the distribution of each of the null statistics $T(\Dxtyz^{(m)})$.
Here we provide assumptions on the data distribution that will ensure that \gls{diet} is able to distinguish between the distribution of the test statistic versus the null statistics.

\begin{restatable}{theorem}{dietpowertheorem}
Let $F_{\cdot\mid \rvz}(\cdot \mid \rvz)$ denote the conditional \gls{cdf} for the distribution $p(\cdot\mid \rvz)$.
Let $\beps = F_{\rvx \mid \rvz}(\rvx \mid \rvz)$ and $\bdelta = F_{\rvy \mid \rvz}(\rvy \mid \rvz)$ be random variables defined over $(\rvx, \rvz)$ and $(\rvy, \rvz)$ respectively.

Assume $F$ is invertible in the first argument and $(\beps, \bdelta) \CI \rvz$.
	If there exists a marginal independence test $\psi: (\sR \times \sR)^N \times [0, 1] \to \{ 0, 1 \}$ that uses a measure of dependence $\rho$ and achieves power greater than $\alpha \in [0, 1]$, then \gls{diet} equipped with $\rho$ and the conditional \glspl{cdf} $F(\cdot \mid \rvz)$ is a conditional independence test with power greater than $\alpha$ for data drawn from $p(\rvx, \rvy, \rvz)$.
\label{thm:diet-cdf-power}	
\end{restatable}

The conditional \glspl{cdf} being invertible is a common assumption: e.g. when $\rvx \sim \mathcal{N}(\rvz_1, \sigma^2)$ or other continuous distributions.
The core assumption here is that $\beps$ and $\bdelta$ are jointly independent of the conditioning set of covariates $\rvz$.
This independence $(\beps, \bdelta) \CI \rvz $ holds in data generating processes where $\rvx, \rvy$ are strictly monotonic transformations of continuous noise variables for any fixed value fo $\rvz$; e.g. additive transformations like $\rvx = \rvz + \text{noise}$ and multiplicative transformations like $\rvx = \rvz*\text{noise}$.
\Cref{appsec:example} shows this formally.

We prove \cref{thm:diet-cdf-power} in \cref{sec:proof-of-diet-cdf-power}: we show that given these conditions, \gls{diet} will provably be able to distinguish between the test and null statistics and achieve power to reject the null hypothesis.
The proof establishes that when $\rvx \not \CI \rvy \mid \rvz$, random variables
$\beps$ and $\bdelta$ will be dependent.
It also shows that under the null hypothesis $\gH_0$, $\beps \CI \bdelta$.
Therefore, the test statistic, which measures the dependence of $\beps$ and $\bdelta$ will have a different distribution than the null statistics.

\paragraph{When is \gls{diet} the most powerful conditionally valid \gls{crt}?}

Here we show that under the same conditions as \cref{thm:diet-cdf-power}, \gls{diet} equipped with a measure of mutual information $\rho$ is the \textit{most powerful} conditionally valid \gls{crt} \citep{katsevich2020theoretical}.

The set of valid \glspl{crt} $\gC_\alpha$ includes any \gls{crt} where the type-1 error is less than $\alpha$ using a dataset $\Dxyz$.
Given samples of $(\rvy, \rvz)$, the set of conditionally valid \glspl{crt} at level $\alpha$ is a subset of $\gC_\alpha$ where the samples of $(\rvy, \rvz)$ in $\Dxyz$ are fixed.
A conditionally valid \gls{crt} is also a marginally valid \gls{crt}.
The following proposition states that given access to the conditional \glspl{cdf} $F_{\rvx \mid \rvz}(\rvx \mid \rvz)$ and $F_{\rvy \mid \rvz}(\rvy \mid \rvz)$, \gls{diet} is the most powerful conditionally valid \gls{crt}.
Thus, the power of \gls{diet} is tied directly to the quality of the estimation of these conditional \glspl{cdf}.
\begin{restatable}{prop}{dietumpprop}
Let 
$\beps = F_{\rvx \mid \rvz}(\rvx \mid \rvz)$
and
$\bdelta = F_{\rvy \mid \rvz}(\rvy \mid \rvz)$.
For data generating processes where both $F_{\cdot\mid \rvz}(\cdot \mid \rvz)$ functions are invertible in the first argument and $(\beps, \bdelta) \CI \rvz$, \gls{diet} with the following mutual information-based marginal dependence measure 
$\rho$ is the most powerful conditionally valid test:
\begin{align*}
\rho(\mathcal{D}_{\bdelta, \beps}) = \frac{1}{N} \sum_{i=1}^N \log \frac{p(\bdelta_i, \beps_i)}{p(\bdelta_i) p(\beps_i)}.	
\end{align*}
\label{prop:ump-info-residuals}
\vspace{-0.35cm}
\end{restatable}

We prove \cref{prop:ump-info-residuals} in \cref{sec:proof-ump-info-residuals} by showing that the likelihood ratio in \cref{prop:ump-info-residuals} is equivalent to the likelihood ratio of $p(\rvy \mid \rvx, \rvz)$ and $p(\rvy \mid \rvz)$: the most powerful conditionally valid \gls{crt} test statistic.

\subsection{Multiple testing and variable selection}

A common application of \glspl{crt} is \acrlong{cvs} \citep{candes2018panning}.
Let $\rvx = \{ \rvx_1, \dots, \rvx_d \}$ be a set of covariates, and $\rvy$ be a response.
\Acrlong{cvs} methods identify a subset of important covariates by testing the conditional independence of each covariate $\rvx_j$ and $\rvy$ given all other covariates $\rvx_{-j}$.
If the hypothesis test for $\rvx_j$ results in a rejection, that variable is ``selected.''
The goal of \acrlong{cvs} is to select as many variables as possible, while controlling for the \gls{fdr}: an analog for type-1 error in multiple testing.

We apply the following procedure to use \gls{diet} for \gls{cvs}.
To test $\rvx_j \CI \rvy \mid \rvx_{-j}$ for each $\rvx_j$, we run \cref{alg:diet} where $\rvz \gets \rvx_{-j}$, $\rvy \gets \rvy$, and $\rvx \gets \rvx_j$.
The resulting set of $p$-values is used with
standard \gls{fdr}-controlling procedures \citep{benjamini1995controlling,benjamini2001control} to select important covariates.

\subsection{Can we further generalize the assumptions made by \gls{diet}?}

Is it possible to generalize the set of distributions for which power is achievable beyond \gls{diet}?
We first outline what a general distillation procedure looks like using functions $u(\rvx,\rvz)$ and $v(\rvy,\rvz)$ to test for conditional independence $\rvx \CI \rvy \g \rvz$.
If these functions $u,v$ are to be learned using samples from $p(\rvx \g \rvz)p(\rvy, \rvz)$ in order to provide type-I error control, we show the challenge faced by a general distillation procedure in always achieving power.

\paragraph{Limits of general distillation procedures.}

Let $L_{\rvx, \rvz}^2$ denote the space of real-valued functions $u$ of $(\rvx, \rvz)$, where $\E [ u(\rvx, \rvz)^2] < \infty$.
Let $L_{\rvy, \rvz}^2$ be defined analogously.
Rather than testing the marginal independence of conditional \glspl{cdf} $F_{\rvx \mid \rvz}(\rvx \mid \rvz)$ and $F_{\rvy \mid \rvz}(\rvy \mid \rvz)$ like \gls{diet}, a \textit{general distillation procedure} tests the marginal independence of some functions  $u \in L_{\rvx, \rvz}^2$ and $v \in L_{\rvy, \rvz}^2$ instead.
\citet{daudin1980partial} shows that,
for all functions $u \in L_{\rvx, \rvz}^2$ and $v \in L_{\rvy, \rvz}^2$ such that $\E[ u(\rvx, \rvz) \mid \rvz ] = 0$ and $\E[ v(\rvy, \rvz) \mid \rvz ] = 0$,
\begin{align*}
    \rvx \CI \rvy \mid \rvz
    \Longleftrightarrow
    \E [u(\rvx, \rvz) v(\rvy, \rvz)] = 0.
\end{align*}
This means that if $\rvy$ is conditionally \textit{dependent} on $\rvx$ given $\rvz$, then there must exist functions $u$ and $v$ such that their correlation is non-zero.
If these $u$ and $v$ are known beforehand, testing their marginal independence will yield a conditional independence test with power.

However, in reality $u$ and $v$ must be learned using data from the data distribution $p(\rvy, \rvx, \rvz)$.
As we show in \cref{corr:error-control}, using data from the \textit{null-data-distribution}\footnote{We use this name to denote that the null dataset $\Dxtyz$, like in \cref{alg:diet}, is sampled from $p_{null}$}, $q_{null} = p(\rvx \g \rvz)p(\rvz, \rvy)$, to learn $u$ and $v$, guarantees \mbox{type-I} error control without the need to sample split or assume the functional form of $\rvy\g \rvx, \rvz$, both of which lead to loss in power.

Learning from the null-data-distribution makes it hard to always achieve power. Consider the following data generating processes:
\begin{align*}
    p_1(\rvy,\rvx,\rvz) & : \\
     \rvy &= \rvx + \rvz \mod 1 \qquad \rvx, \rvz \sim \text{Uniform}(0, 1)  \\
    p_2(\rvy,\rvx,\rvz) & : \\
	    \rvy &= \rvx \qquad \rvx, \rvz \sim \text{Uniform}(0,1),
\end{align*}
where $\rva + \rvb \mod 1$ is defined as $\rva + \rvb$ if $\rva + \rvb < 1$ and $\rva + \rvb - 1$ if $\rva + \rvb \geq 1$.
Note that the marginals of $(\rvx, \rvz)$ and $(\rvy, \rvz)$ are the same across both $p_1$ and $p_2$.
In turn, the null-data-distributions are equal, $p_1(\rvx \g \rvz)p_1(\rvz, \rvy) = p_2(\rvx \g \rvz)p_2(\rvz, \rvy)$, meaning that any distillation procedure will learn the same functions $u,v$ in either distribution. However, the same $u,v$ can have dramatically different power in $p_1$ and $p_2$ making it difficult to build a generic distillation procedure.
For example, let $u(\rvx, \rvz) = \rvx - 0.5$ and $v(\rvy, \rvz) = \rvy - 0.5$.
Any general distillation procedure that tests the correlation between these $u,v$ would yield power in $p_2$
but would have no power under $p_1$ because $u,v$ are independent under $p_1$.

\paragraph{When do distillation procedures achieve power?}

Let $\beps=u(\rvx, \rvz), \bdelta=v(\rvy, \rvz)$ be the variables computed by a distillation procedure, like \gls{diet} or  $d_0$-\gls{crt}.
The previous subsection explains the challenge any distillation procedure faces in both achieving power and having type-I error control.
Here, we give conditions on the data generating process and the variables $\beps, \bdelta$ computed by the distillation procedure that guarantee power.

\begin{theorem}
Consider any data generating process of the following form:
\[\rvz \sim p(\rvz), \quad \rve, \rvd \sim p(\rve, \rvd), \quad \rvx = f(\rve,\rvz) \quad \rvy = g(\rvd,\rvz).\]
Let $\beps, \bdelta$ be distributed according to:
\[(\beps, \bdelta, \rvx, \rvy, \rvz) \sim \qhat(\beps, \bdelta \mid \rvx, \rvy, \rvz) p(\rvx, \rvy, \rvz).
\]

Further let,
\begin{align}
	\qhat&(\beps, \bdelta \mid \rvx, \rvy, \rvz) = p(\beps \mid \rvx, \rvz) p(\bdelta \mid \rvy, \rvz), \tag{factorization} \label{eqn:factorization} \\
	\exists &\tilde{f}, \tilde{g} \quad 
	\quad \rvx \overset{\text{a.s.}}{=} \tilde{f}(\beps, \rvz), \quad \rvy \overset{\text{a.s.}}{=} \tilde{g}(\bdelta, \rvz), \tag{reconstruction} \label{eqn:reconstruction} \\
	(&\rvd, \bdelta) \CI \rvz \qquad (\rve, \beps) \CI \rvz.
	 \tag{joint independence} \label{eqn:joint-independence}
\end{align}
Let $\psi(\Dbde, \alpha): (\sR \times \sR)^N \times [0,1] \to \{0, 1\}$ be a marginal independence test that uses statistic $\rho: (\sR \times \sR)^N \to \sR$ and has power greater than $\alpha$.
Let $\Dbde$ be a dataset of $N$ samples of $(\beps, \bdelta)$ generated using $\qhat(\beps, \bdelta \mid \rvx, \rvy, \rvz)$ and $\Dxyz$.
Then, $\psi$ using $\Dbde$ and $\rho$ is also a conditional test of independence for $\rvx \CI \rvy \mid \rvz$ with power greater than $\alpha$.
\label{thm:diet-general-main}
\end{theorem}

We prove \cref{thm:diet-general-main} in \cref{sec:proof-of-sufficient-conditions-for-power}.
\Cref{thm:diet-general-main} allows one to use knowledge about the form of the data generating process to understand whether a distillation procedure achieves power.

As an example, see \cref{sec:dcrt-fits-into-our-theorem} where we show how the $d_0$-\gls{crt} satisfies the conditions in \cref{thm:diet-general-main} for additive data generating processes and therefore achieves power for such processes.

%% file: sections/experiments.tex
We analyze the performance of \acrshort{diet}  on several synthetic and real datasets and compare it to well-studied methods designed to make \glspl{crt} tractable.

\paragraph{\Gls{diet} setup.}

The \glspl{mdn} in \gls{diet} take $\rvz$ as input and use a six-layer fully-connected network with batch normalization and ReLU activations to output the parameters of a Gaussian mixture with 10 components.
As a marginal dependence statistic $\rho$, we use the 
 mutual information estimator from \citet{vinh2009information}.
Further training and hyperparameter details are given in \cref{sec:diet-mdn-details}.

\paragraph{Baselines.}

We use the $d_0$-\gls{crt} and $d_I$-\gls{crt} models described by \citet{liu2020fast}.
The top-$k$ $\rvz$ dimensions are chosen using the Lasso heuristic proposed by \citet{liu2020fast}.
This model regresses $\rvy$ onto $\rvz \in \sR^p$ and picks the top $k = 2 \log p$ dimensions of $\rvz$ with the largest absolute regression coefficients.

The \glspl{hrt} we include in our experiments use a model $\qmodel(\rvy \mid \rvx, \rvz)$ that consists of a six-layer fully-connected network with batch normalization and ReLU activations.
We implement the cross-validated version of the \gls{hrt} suggested by \citet{tansey2018holdout} that achieves higher power in finite samples.

Further details like the test statistics used for each baseline method can be found in \cref{sec:baseline-details}.

\paragraph{Experiment details.}

Each synthetic experiment follows the same basic structure for a single run, unless specified otherwise.
First, a dataset $\Dxyz$ is sampled.
Then, each method is used to test the hypothesis $\rvx \CI \rvy \mid \rvz$ and a $p$-value is computed using $M=100$ null datasets.
We perform 100 runs of each synthetic experiment and report aggregate results.

The power of each method at a specific rejection threshold $\alpha$ is estimated by computing the percentage of times a hypothesis is rejected, over the 100 runs.
A hypothesis is rejected if the $p$-value $\pval \leq \alpha$.

For \acrlong{cvs} experiments, we test the hypothesis $\rvx_j \CI \rvy \mid \rvx_{-j}$ for each dimension $j$ of the covariate vector $\rvx$.
We then apply the Benjamini-Hochberg procedure \citep{benjamini1995controlling} to  account for multiple testing while controlling the \gls{fdr}.

To test each method in a realistic setting, the \acrlong{cvs} experiments use only a fixed set of $\rvx$ samples.
To generate the null datasets $\{ \Dxtyz^{(m)} \}_{m=1}^M$, we employ a deep generative model to jointly model each $p(\rvx_j \mid \rvx_{-j})$ distribution \citep{romano2020deep}.
\Cref{sec:cvs-details} provides an overview of this process.
Since a deep generative model must be fit to generate null datasets,
this experiment uses half the available data to fit the model, while the other half is used to run each \gls{crt}.
Each synthetic variable selection experiment is run 100 times.
We set $M = 2000$.

\subsection{Synthetic experiments}

\paragraph{Univariate Gaussian data.}

This experiment is designed mainly to confirm that each method performs as intended.
The data is drawn as follows: $\rvz \sim \mathcal{N}(0, 0.1)$, $\rvx \mid \rvz \sim \mathcal{N}(\rvz, 0.1)$, and $\rvy \mid \rvx, \rvz \sim \mathcal{N}(\rvx + \rvz, 0.1)$.
The training dataset consists of 500 samples.

\textit{Results:} As expected, the estimated power of each method is 1 for $\alpha \in (0, 0.3]$.
We do not explore larger $\alpha$, as a practitioner would realistically set their nominal error rate within this range.
As a graph is unnecessary to visualize this result, we omit it.

\paragraph{Non-Gaussian and multiplicative data.}

These experiments are designed primarily to understand the effect of violating an additivity assumption in the data generating process.
Using noise $\varepsilon \sim \mathcal{N}(0, 0.01)$ and coefficients $\beta \in \mathbb{R}^{100}$ where each $\beta_j \sim \mathcal{N}(0, 1)$ and sorted so that 
$|\beta_1| \geq |\beta_2| \geq \cdots, \geq \beta_d$, and
 $ \rvz  \sim \mathcal{N}(0, 0.01 \cdot I_{100})$
\begin{align}
	\rvx & \mid \rvz \sim \mathcal{N}\left(\sum_{j=1}^{10} \beta_j \rvz_j, 0.25 \right) \nonumber \\
	\textstyle \rvy & \mid \rvx, \rvz, \varepsilon = (\rvx + \varepsilon + \sum_{j=1}^{100} \rvz_j \beta_j)^3 \tag{Non-Gaussian} \\
	\rvx & \sim \mathcal{N}(0, 1) \nonumber \\
	\rvy & \mid \rvz, \rvx, \epsilon = 4\beta_1\rvz_1 \rvx + 4 \beta_2 \rvz_2 + \epsilon \tag{Multiplicative}
\end{align}
Both datasets consist of 1000 samples.

\setlength{\columnsep}{5pt}
\begin{figure}[t]
\centering\includegraphics[width=\linewidth]{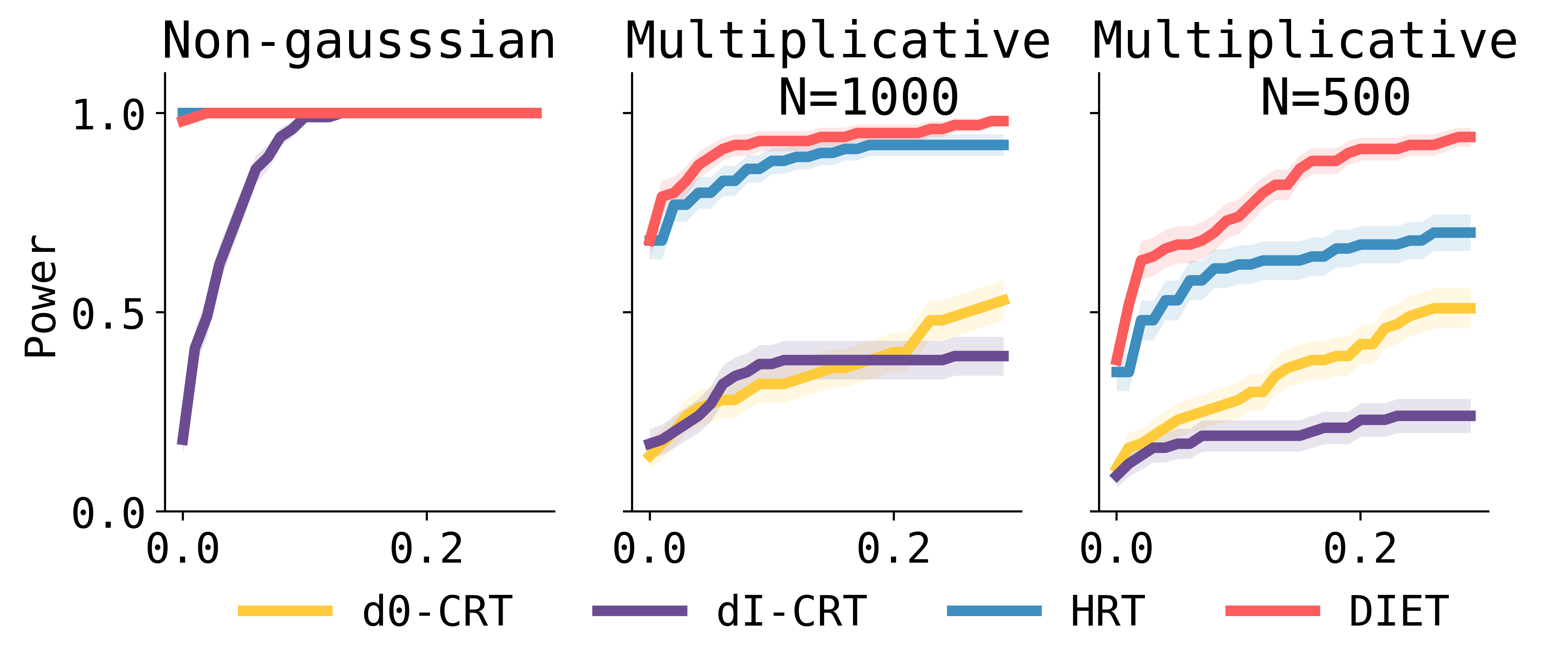}
  \caption{\textbf{\Gls{diet} achieves high power across numerous synthetic benchmarks.} In this figure, we show the power of each method as a function of nominal type-1 error rate $\alpha$.}\label{fig:power-plot}
\end{figure}

\textit{Results:} We observe that each \gls{crt} manages to control the type-1 error rate at or below nominal levels.
In terms of power, most methods perform well on the non-Gaussian dataset, as shown in the first column of \cref{fig:power-plot}.
All but the $d_I$-\gls{crt} are able to achieve full power for almost every $\alpha \in (0, 0.3]$.

\setlength{\belowcaptionskip}{-10pt}
\begin{figure}
  \centering\includegraphics[width=0.5\linewidth]{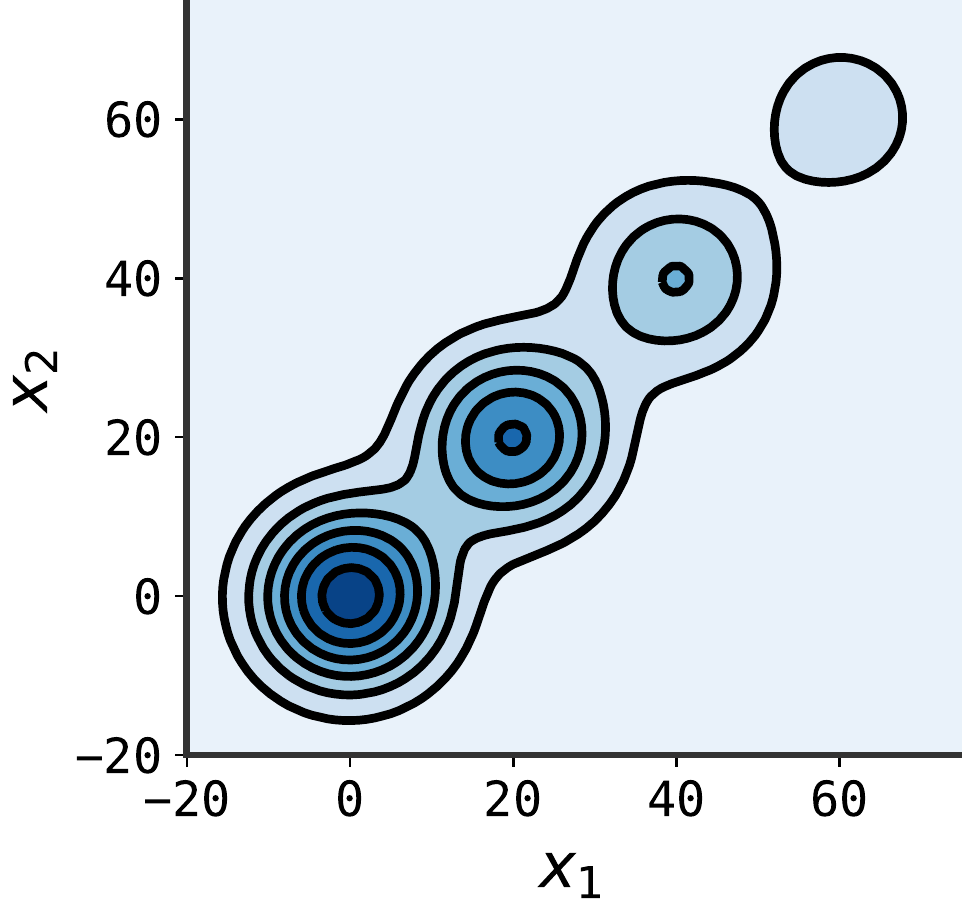}
  \caption{\textbf{Synthetic \gls{cvs} dataset}}
  \label{fig:mix}
\end{figure}

In the case of multiplicative data, there is a clear deterioration in the performance of the $d_0$-\gls{crt} and $d_I$-\gls{crt}, as shown in the second column of \cref{fig:power-plot}.
The $d_I$-\gls{crt} achieves marginally higher power for $\alpha < 0.2$, but is still quite far from \acrshort{diet} or \gls{hrt}.
Upon investigation, we observed that the heuristic used to choose dimensions in $\rvz$ in $d_I$-\gls{crt} only selects $\rvz_1$ at random.
Since \glspl{dcrt} forbid using samples of the triple $(\rvx, \rvy, \rvz)$ during training, it is difficult to choose a robust heuristic.
We explore why \gls{diet} achieves higher power from a theoretical perspective in \cref{sec:dIcrt-counterexample}.

Then, to understand the cost of sample splitting, we reduced the sample size of the multiplicative data to $500$ and re-ran our experiments.
The third column of \cref{fig:power-plot} shows that the \gls{hrt} suffers the greatest loss in power.
This is likely due to the \gls{hrt} splitting the sample and using only $200$ samples during training.

\paragraph{Controlled variable selection.}

This experiment evaluates each \gls{crt} on its ability to perform controlled variable selection while using an estimated $p(\rvx \mid \rvz)$ distribution.
The $\rvx$ is a 100-dimensional mixture of autoregressive Gaussians.
\Cref{fig:mix} visualizes the first two dimensions of this data.
The response $\rvy \mid \rvx$ is a conditional Gaussian whose mean is a linear function of $\rvx$ with only 20 non-zero coefficients.
We refer the reader to \cref{sec:synthetic-cvs-details} for the exact sampling process.
The dataset consists of $1000$ samples.

\begin{figure}
\centering\includegraphics[width=0.65\linewidth]{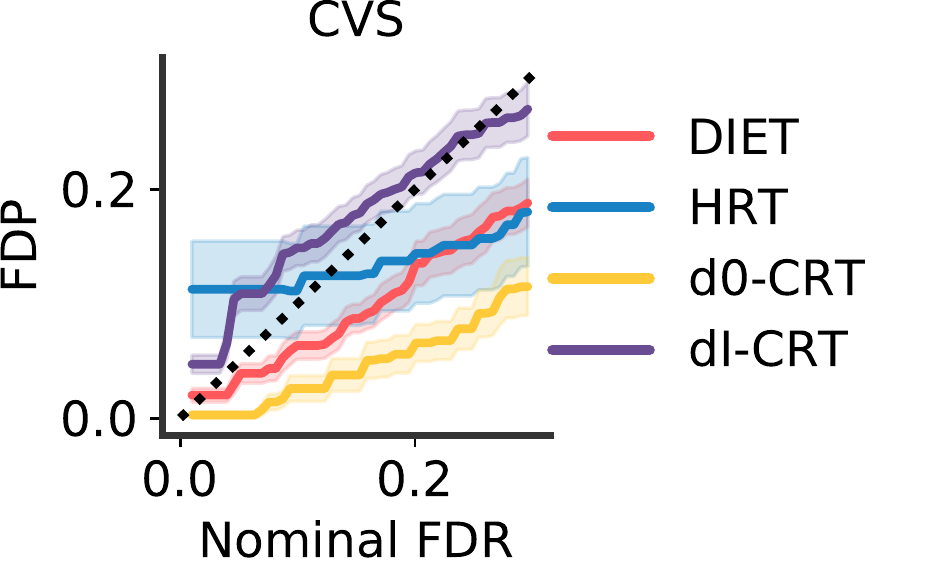}
  \caption{\textbf{\Gls{fdp} of each method on synthetic \gls{cvs} data.}}
  \label{fig:cvs-fdp}
\end{figure}

\textit{Results:} We evaluate the average power and the \gls{fdp} across runs for each method in the fourth column of \cref{fig:power-plot-cvs} and \cref{fig:cvs-fdp} respectively.
The average \gls{fdp} is an empirical estimate of the \gls{fdr}.
We notice that most methods are able to keep the average \gls{fdp} below the nominal \gls{fdr} rate $\alpha$ for $\alpha > 0.2$.
However, when $\alpha \leq 0.1$, the $d_I$-\gls{crt} and the \gls{hrt} inflate the \gls{fdp}, suggesting they are sensitive to poor estimations of the $p(\rvx_j \mid \rvx_{-j})$ distributions, as shown by \citet{sudarshan2021contra}.
We also observe that loss of power in the \gls{hrt} is mainly due to sample splitting.
Using 3000 samples instead helped increase the power of the \gls{hrt} closer to that of \gls{diet}.

\subsection{Semi-synthetic genetics experiment}

A common application area of Model-X methods is biology \citep{candes2018panning,bates2020causal,sudarshan2020deep,sesia2019gene}.
We evaluate each \gls{crt} using a setup similar to that of \citet{sudarshan2020deep}, which uses RNA expression data of 963 cancer cell lines and 20K genes per cell line from \citet{yang2012genomics}.
The datasets $\Dxyz \in \sR^{963 \times 100}$ are generated as follows.

100 genes are sampled sequentially from the set of 20K such that the resulting set contains genes with strong pairwise correlations.
We use a synthetic $\rvy \mid \rvx$ response function 
from \citet{tansey2018holdout}.
\Cref{sec:semisynthetic-cvs-details} contains specific details about the dataset creation.
We perform 30 replicates of this experiment; $\rvx_{1:20}$ are the important features in each one.

\begin{figure}[t]
\centering\includegraphics[width=1.0\linewidth]{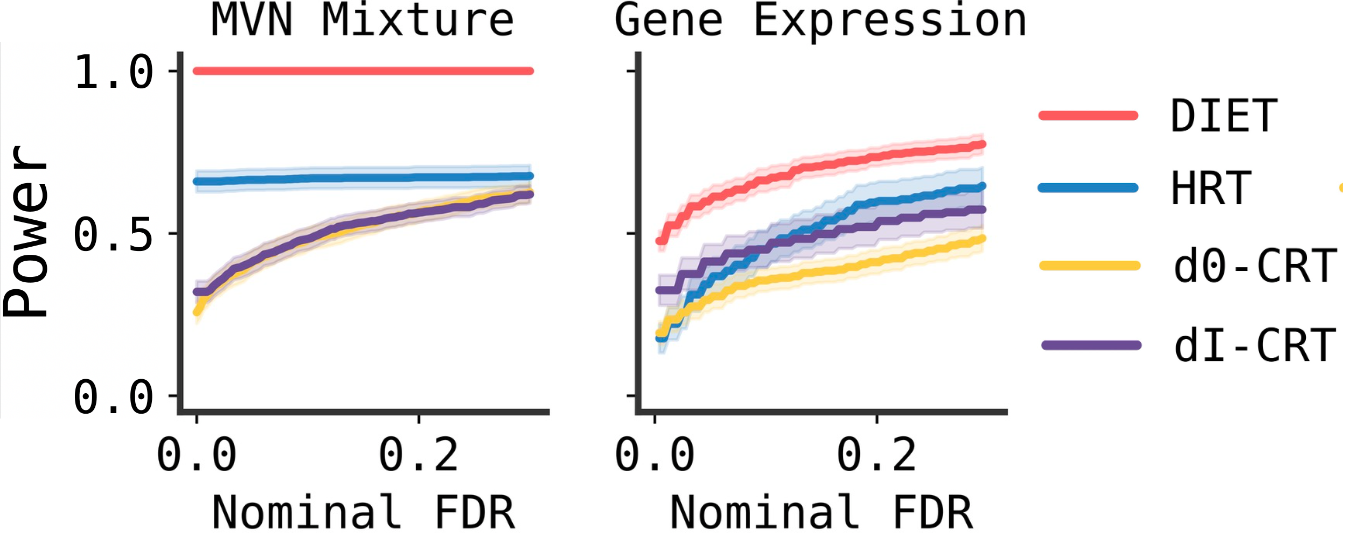}
  \caption{\textbf{\Gls{diet} achieves high power in CVS experiments.}
  In this figure, we show the average power over 100 repetitions of each method as a function of nominal  \gls{fdr} in the case of variable selection.}\label{fig:power-plot-cvs}
\end{figure}

\textit{Results:} We show the average power for each \gls{crt} in the last column of \cref{fig:power-plot-cvs}.
All methods are able to control the average \gls{fdp} below the nominal level.
\Acrshort{diet} consistently achieves power higher than the baselines.
We also observe that the \gls{hrt} achieves higher power than the $d_I$-\gls{crt} at nominal \gls{fdr} above 0.1.
At lower nominal \gls{fdr}, the \gls{hrt} does not select many features as its non-null $p$-values are generally higher than those of the $d_I$-\gls{crt}.

\setlength{\belowcaptionskip}{-10pt}

\subsection{Electronic health records}

\Glspl{crt} have found use in clinical model deployment pipelines as methods to prune a set of input features \citep{razavian2020validated}.
This pruning reduces the amount of auditing and engineering needed for model deployment.
We perform \acrlong{cvs} using an \gls{ehr} dataset from a large metropolitan hospital to understand which variables are most predictive of an adverse event within 96 hours for patients that tested positive for COVID-19.

The data contains 28K samples with 29 features on the results of a blood test, basic vital signs, and demographics.
A full list of variables is provided in \cref{sec:additional-figures}.
We run each \gls{crt} method on the \gls{ehr} dataset and apply the \citet{benjamini1995controlling} procedure, selecting covariates at a nominal \gls{fdr} of $10\%$.

\begin{table}[t]
\vspace{20pt}
\centering
\caption{\small{\textbf{\Acrshort{diet} selects a larger portion of covariates previously identified by highly-cited medical papers.} See \cref{sec:additional-figures} for a list of selections.}}
\label{tab:covid-selections}
\begin{tabular}{@{}lcccc@{}}
\toprule
                 & \acrshort{diet}  & \gls{hrt}  & $d_0$-\gls{crt}   & $d_I$-\gls{crt}   \\ \midrule
Selected & \textbf{60\%} & 40\% & 25\% & 55\% \\ \bottomrule
\end{tabular}
\end{table}

\textit{Results:} 
To evaluate the effectiveness of the selections made by each \gls{crt}, we compare selected covariates to those reported by several papers related to adverse events in COVID-19 patients from well-known medical journals
\citep{petrilli2020factors,sattar2020obesity,mei2020risk,castro2020laboratory,zhang2020d,zhong2021mean,ruan2020clinical,zhou2020clinical}.

To score each \gls{crt}, we consider covariates found to be important by at least one of the above papers.
We compute the fraction of these covariates selected by each \gls{crt} and report them in \cref{tab:covid-selections}.
We show the full list of selections in \cref{sec:additional-figures}.

\Acrshort{diet} selects a larger percent of the important covariates, which indicates higher power.
While the $d_I$-\gls{crt} selects almost as many, 
upon closer inspection, it also selects redundant features.
For example, the $d_I$-\gls{crt} selects both count and percentage of Eosinophils, and both High O2 support and O2 device while \acrshort{diet} only selects one of each.

%% file: sections/discussion.tex
Existing methods to speed up model-based \glspl{crt} either make restrictive assumptions about the data generating process, use heuristics to model interactions between $\rvx$ and $\rvy$, or lose power due to sample splitting.
\Gls{diet} provides a flexible way to avoid each of these issues and is applicable to a wide range of data generating distributions.
It uses conditional \gls{cdf} estimators to reduce high-dimensional model-based \glspl{crt} to tests of marginal independence. 

We show theoretically that \gls{diet} will achieve type-1 error control regardless of data distribution $p(\rvx, \rvy, \rvz)$, then we characterize a class of data distributions for which \gls{diet} can provably achieve power. 
Future work in this area can study weaker assumptions on the data generating process to provably achieve power in a distillation-based \gls{crt}.
This can lead to further insight into when a conditional independence test can be reduced to a marginal one
 without sacrificing power.

%% file: sections/appendix_proofs.tex
\subsection{Shortcomings of \glspl{dcrt} example}
\label{sec:dIcrt-counterexample}

Consider the following example from earlier:
\begin{align*}
\rvx &\sim \mathcal{N}(\rvx; 0, \sigma_\rvx^2) \\
\rvz_j &\sim \mathcal{N}(\rvz_j; 0, 1) \, \forall j \in \{1, \dots, d \} \\
\rvy \mid \rvx, \rvz &\sim \mathcal{N}(\rvy; \beta_1 \rvx \rvz_1+\sum_{j = 2}^d \beta_j \rvz_j, 1) \\
\end{align*}

Since this example extends the motivating example for $d_I$-\glspl{crt} from \citet{liu2020fast}, we focus only on the behavior of the $d_I$-\gls{crt} here.
Recall the $d_I$-\gls{crt} test statistic computation:
\begin{enumerate}
	\item The $d_I$-\gls{crt} first identifies a subset of $k$ variables in $\rvz$ with which to explicitly compute interaction terms. This is done by fitting a regression from  $\rvz$ to $\rvy$, then using some measure of feature importance to select the top $k$ most important features, $\rvz_{\mathrm{top}(k)}$
	\item The distillation function $d_\rvy = \mathbb{E}[\rvy \mid \rvz]$ is computed
	\item Then, the distillation function $d_\rvx = \mathbb{E}[\rvx \mid \rvz]$ is computed
	\item Next, a model from $(\rvx - d_\rvx, d_\rvy, \rvz_{\mathrm{top}(k)})$ to $\rvy$ is fit
	\item Finally, a measure of feature importance for $\rvx - d_\rvx$ in this model is used to compute the test statistic $T$
\end{enumerate}

To compute each null statistic, steps 3-5 are repeated using the null datasets. Given the set of $M$ null statistics and the test statistic $T$, a $p$-value is computed as shown in the \gls{crt} $p$-value computation \cref{eqn:crt-pvalue}.
Now, observe the behavior of the $d_I$-\gls{crt} in this example.

First, a model is fit from $\rvz$ to $\rvy$. This is equivalent to estimating the function $\mathbb{E}[\rvy \mid \rvz]$. To see the functional form of this quantity let's first evaluate the density $F_{\rvy \mid \rvz}(\rvy \mid \rvz)$:
\begin{align*}
F_{\rvy \mid \rvz}(\rvy \mid \rvz) &= \int_{-\infty}^{\infty} f(\rvy \mid \rvx, \rvz) F_{\rvx \mid \rvz}(\rvx \mid \rvz) d\rvx \\
&=  \int_{-\infty}^{\infty} f(\rvy \mid \rvx, \rvz) f(\rvx) d\rvx \\
&= \int_{-\infty}^{\infty} \frac{\mathrm{e}^{-\frac{\left(-{\beta_1}\rvz_1\rvx-{\sum_{j=2}^d \beta_j \rvz_j}+\rvy\right)^2}{2}-\frac{\rvx^2}{2{\sigma_\rvx}^2}}}{\sqrt{4{\pi}^2{\sigma_\rvx}^2}}\,\mathrm{d}\rvx \\
&= \int_{-\infty}^{\infty} \frac{\mathrm{e}^{-\frac{\left(-\ZT\rvx+\ZY\right)^2}{2}-\frac{\rvx^2}{2{\sigma_\rvx}^2}}}{\sqrt{4{\pi}^2{\sigma_\rvx}^2}}\,\mathrm{d}\rvx \qquad \{\text{letting } \ZT = \beta_1 \rvz_1, \ZY = \rvy - \sum_{j=2}^d \beta_j \rvz_j\}
\\
    &= \int_{-\infty}^{\infty} \frac{\mathrm{e}^{-\frac{\ZT^2\rvx^2- 2\ZY\ZT \rvx + \ZY^2}{2}-\frac{\rvx^2}{2{\sigma_\rvx}^2}}}{\sqrt{4{\pi}^2{\sigma_\rvx}^2}}\,\mathrm{d}\rvx
\\
    &= \frac{\mathrm{e}^{-\nicefrac{\ZY^2}{2}}}{\sqrt{4{\pi}^2{\sigma_\rvx}^2}} \int_{-\infty}^{\infty}  \mathrm{e}^{-\frac{\ZT^2\rvx^2- 2\ZY\ZT \rvx  }{2}-\frac{\rvx^2}{2{\sigma_\rvx}^2}}\,\mathrm{d}\rvx
\\
    &= \frac{\mathrm{e}^{-\nicefrac{\ZY^2}{2}}}{\sqrt{4{\pi}^2{\sigma_\rvx}^2}} 
    \int_{-\infty}^{\infty}  
        \mathrm{e}^{-\frac{1}{2}\left(\rvx^2 \left(\ZT^2  + \frac{1}{{\sigma_\rvx}^2} \right) 
        {- 2\ZY\ZT \rvx +  \frac{\ZY^2 \ZT^2}{\left(\ZT^2  + \frac{1}{{\sigma_\rvx}^2} \right) } - \frac{\ZY^2 \ZT^2}{\left(\ZT^2  + \frac{1}{{\sigma_\rvx}^2} \right) } } \right)}\,  
    \mathrm{d}\rvx
\\
    &= \frac{\mathrm{e}^{-\nicefrac{\ZY^2}{2}}}{\sqrt{4{\pi}^2{\sigma_\rvx}^2}} 
    \int_{-\infty}^{\infty}  
        \mathrm{e}^{-\frac{1}{2}\left(\rvx\left(\ZT^2  + \frac{1}{{\sigma_\rvx}^2} \right)^{\nicefrac{1}{2}}
         - { \frac{\ZY \ZT}{\left(\ZT^2  + \frac{1}{{\sigma_\rvx}^2} \right)^{\nicefrac{1}{2}}}} \right)^2  + \frac{\ZY^2 \ZT^2}{\left(\ZT^2  + \frac{1}{{\sigma_\rvx}^2} \right) }}\,
    \mathrm{d}\rvx
\\
    &= \frac{\mathrm{e}^{-\nicefrac{\ZY^2}{2}  + \frac{\ZY^2 \ZT^2}{2\left(\ZT^2  + \frac{1}{{\sigma_\rvx}^2} \right) }}}{\sqrt{4{\pi}^2{\sigma_\rvx}^2}} 
    \int_{-\infty}^{\infty} 
        \mathrm{e}^{-\frac{\left(\ZT^2  + \frac{1}{{\sigma_\rvx}^2} \right)}{2}\left(\rvx
         - { \frac{\ZY \ZT}{\left(\ZT^2  + \frac{1}{{\sigma_\rvx}^2} \right)}} \right)^2 }\,
    \mathrm{d}\rvx
\\
    &= \frac{\mathrm{e}^{-\ZY^2\left(\frac{1}{2}  - \frac{1 }{2\left(1  + \frac{1}{{\ZT^2\sigma_\rvx}^2} \right) }\right)
    }}{\sqrt{4{\pi}^2{\sigma_\rvx}^2}} 
    \int_{-\infty}^{\infty} 
        \mathrm{e}^{-\frac{\left(\ZT^2  + \frac{1}{{\sigma_\rvx}^2} \right)}{2}\left(\rvx
         - { \frac{\ZY \ZT}{\left(\ZT^2  + \frac{1}{{\sigma_\rvx}^2} \right)}} \right)^2 }\,
    \mathrm{d}\rvx
\\
    &= \frac{\mathrm{e}^{-\ZY^2\left(\frac{1}{2}  - \frac{1 }{2\left(1  + \frac{1}{{\ZT^2\sigma_\rvx}^2} \right) }\right)
    }}{\sqrt{4{\pi}^2{\sigma_\rvx}^2}} 
    \sqrt{\frac{2\pi}{\left(\ZT^2  + \frac{1}{{\sigma_\rvx}^2} \right)}}
\\
    &= \frac{\mathrm{e}^{-\ZY^2\left(\frac{\left(1  + \frac{1}{{\ZT^2\sigma_\rvx}^2} \right) - 1}{2\left(1  + \frac{1}{{\ZT^2\sigma_\rvx}^2} \right) }\right)
    }}{\sqrt{2{\pi}\left({\sigma_\rvx}^2\ZT^2  + 1 \right)}}
\\
    &= \frac{\mathrm{e}^{-\frac{\ZY^2}{2\left({\sigma_\rvx}^2\ZT^2  + 1\right)}
    }}{\sqrt{2{\pi}\left({\sigma_\rvx}^2\ZT^2  + 1 \right)}}
\\
    &= \frac{\mathrm{e}^{-\frac{\left(\rvy - \sum_{j=2}^d \beta_j \rvz_j\right)^2}{2\left({\sigma_\rvx}^2\ZT^2  + 1\right)}
    }}{\sqrt{2{\pi}\left({\sigma_\rvx}^2\ZT^2  + 1 \right)}}
\\
&= \mathcal{N}(\rvy;\sum_{j=2}^d \beta_j \rvz_j, 1+\beta_1^2\sigma_\rvx^2\rvz_1^2).
\end{align*}

This is a Gaussian distribution with mean $\mathbb{E}[\rvy \mid \rvz]=\sum_{j=2}^d \beta_j \rvz_j$, which is not a function of $\rvz_1$. Therefore, $\rvz_{\mathrm{top}(k)}$ will not include $\rvz_1$ for any $k < d$. To compute $d_\rvx=\mathbb{E}[\rvx \mid \rvz]$, note that $\rvx$ and $\rvz$ are independent, and $\mathbb{E}[\rvx] = 0$.

Next, let's consider a model from $(\rvx - d_\rvx, d_\rvy, \rvz_{\mathrm{top}(k)})$ to $\rvy$. Again, this is equivalent to estimating $\mathbb{E}[\rvy \mid \rvx - d_\rvx, d_\rvy, \rvz_{\mathrm{top}(k)}] = \mathbb{E}[\rvy \mid \rvx, \sum_{j=2}^d \beta_j \rvz_j, \rvz_{\mathrm{top}(k)}]$. Since $\rvz_1$ is not in the conditioning set of this expectation, it reduces to $\mathbb{E}[\rvy \mid \sum_{j=2}^d \beta_j \rvz_j, \rvz_{\mathrm{top}(k)}]$; this follows from expanding the conditional expectation and noting $\E[\rvz_1] = 0$. Thus any model from $(\rvx - d_\rvx, d_\rvy, \rvz_{\mathrm{top}(k)})$ to $\rvy$ will assign no feature importance to $\rvx - d_\rvx$. Assuming that a feature importance score of 0 indicates an unimportant feature, the score assigned to $\rvx - d_\rvx$ will be 0.

The same holds true when repeating the $d_I$-\gls{crt} steps 3-5 with the null datasets. Regardless of what values of $\rvx$ are used in the model that estimates $\mathbb{E}[\rvy \mid \sum_{j=2}^d \beta_j \rvz_j, \rvz_{\mathrm{top}(k)}]$, the importance score of $\rvx - d_\rvx$ will always be zero. Since the distribution of the test statistic is indistinguishable from the distribution of the null statistics, the $d_I$-\gls{crt} will achieve power no greater than the size of the test.

Next, consider the case of \gls{diet}.
Recall that its test statistic uses the dataset $D_{\rvx,\rvy,\rvz} = \{ (\rvx^{(i)}, \rvy^{(i)}, \rvz^{(i)}) \}_{i=1}^n$ to compute samples of $\bdelta = F_{\rvy \mid \rvz}(\rvy, \rvz)$ and $\beps = F_{\rvx \mid \rvz}(\rvx, \rvz) = F_\rvx(\rvx)$, then uses these samples to estimate the marginal dependence between $\bdelta$ and $\beps$.
We will now show that in the example above, $\bdelta$ and $\beps$ will be dependent using the true data $D_{\rvx,\rvy,\rvz}$, but will be independent when using the null data $D_{\tilde{\rvx},\rvy,\rvz}$, yielding power $>0$.

First note the following equivalences:

\begin{align*}
F_{\rvy \mid \rvz}(\rvy \mid \rvz) &= \Phi\left(\frac{\rvy - \sum_{j=2}^d \beta_j \rvz_j}{\sqrt{1 + \beta_1^2 \sigma_\rvx^2 \rvz_1^2}}\right) \\
F_\rvx(\rvx) &= \Phi\left(\frac{\rvx}{\sigma_\rvx}\right) \\
\rvy &= \beta_1 \rvx \rvz_1 + \sum_{j=2}^d \beta_j \rvz_j + \eta_{\rvy}
\end{align*}

where $\Phi$ is the \gls{cdf} of a standard gaussian and $\eta_{\rvy} \sim \mathcal{N}(0, 1)$. To show that $\bdelta$ and $\beps$ are dependent, we must show that 

\begin{gather*}
\mathbb{P}(\bdelta \leq a \mid \beps = b) \not= \mathbb{P}(\bdelta \leq a).
\end{gather*}

When using the true data $D_{\rvx,\rvy,\rvz}$, the following must hold:
\begin{align*}
\mathbb{P}(\bdelta \leq a \mid \beps = b) &= \mathbb{P}\left(\Phi\left(\frac{\rvy - \sum_{j=2}^d \beta_j \rvz_j}{\sqrt{1 + \beta_1^2 \sigma_\rvx^2 \rvz_1^2}}\right) \leq a \mid \Phi\left(\frac{\rvx}{\sigma_\rvx}\right) = b \right) \\
&= \mathbb{P}\left(\frac{\rvy - \sum_{j=2}^d \beta_j \rvz_j}{\sqrt{1 + \beta_1^2 \sigma_\rvx^2 \rvz_1^2}} \leq \Phi^{-1}(a) \mid \rvx = \sigma_\rvx \Phi^{-1}(b) \right) \\
&= \mathbb{P}\left(\frac{\beta_1 \rvx \rvz_1 + \eta_{\rvy}}{\sqrt{1 + \beta_1^2 \sigma_\rvx^2 \rvz_1^2}} \leq \Phi^{-1}(a) \mid \rvx = \sigma_\rvx \Phi^{-1}(b) \right) \\
&= \mathbb{P}\left(\frac{\beta_1 \rvz_1 \sigma_\rvx \Phi^{-1}(b) + \eta_{\rvy}}{\sqrt{1 + \beta_1^2 \sigma_\rvx^2 \rvz_1^2}} \leq \Phi^{-1}(a) \right).
\end{align*}
The first equation uses the definitions of $\bdelta$ and $\beps$. The second equation uses the invertibility of the Gaussian CDF. The third equation holds because $\rvy$ can be rewritten as a function of $\rvx$, $\rvz$, and noise $\eta_{\rvy}$. Finally, the last equation uses the value of $\rvx$ as a function of $b$ and that $\rvx$ is jointly independent of $\rvz_1$ and $\eta_{\rvy}$. Clearly, the conditional probability $\mathbb{P}(\bdelta \leq a \mid \beps = b)$ cannot be written as $\mathbb{P}(\bdelta \leq a)$ using the true data $D_{\rvx,\rvy,\rvz}$. This means that $\bdelta$ and $\beps$ will be dependent.

When computing the dependence of $\bdelta$ and $\beps$ using null datasets:
\begin{align*}
\mathbb{P}(\bdelta \leq a \mid \beps = b) &= \mathbb{P}\left(\Phi\left(\frac{\rvy - \sum_{j=2}^d \beta_j \rvz_j}{\sqrt{1 + \beta_1^2 \sigma_\rvx^2 \rvz_1^2}}\right) \leq a \mid \Phi\left(\frac{\tilde{\rvx}}{\sigma_\rvx}\right) = b \right) \\
&= \mathbb{P}\left(\frac{\rvy - \sum_{j=2}^d \beta_j \rvz_j}{\sqrt{1 + \beta_1^2 \sigma_\rvx^2 \rvz_1^2}} \leq \Phi^{-1}(a) \mid \tilde{\rvx} = \sigma_\rvx \Phi^{-1}(b) \right) \\
&= \mathbb{P}\left(\frac{\beta_1 \rvx \rvz_1 + \eta_{\rvy}}{\sqrt{1 + \beta_1^2 \sigma_\rvx^2 \rvz_1^2}} \leq \Phi^{-1}(a) \mid \tilde{\rvx} = \sigma_\rvx \Phi^{-1}(b) \right) \\
&= \mathbb{P}\left(\frac{\beta_1 \rvx \rvz_1 + \eta_{\rvy}}{\sqrt{1 + \beta_1^2 \sigma_\rvx^2 \rvz_1^2}} \leq \Phi^{-1}(a) \right) \\
&= \mathbb{P}\left(\bdelta \leq a \right).
\end{align*}
The first 3 equations follow from earlier. The 4th and 5th steps hold because $\rvy$ is not a function of $\tilde{\rvx}$ and $\tilde{\rvx}$ is jointly independent of all other random variables. Therefore, when computing each null statistic using null data $D_{\tilde{\rvx},\rvy,\rvz}$, $\bdelta$ and $\beps$ will be independent.

Since \acrshort{diet-cdf} will identify dependence between $\bdelta$ and $\beps$ when using the true data, and no dependence when using the null data, the distribution of the test statistic will not be equal to that of each null statistic. Thus, it follows that \acrshort{diet-cdf} can achieve power $> 0$.

\subsection{Proof of \cref{prop:type-1-error}}
\label{sec:proof-type-1-error}

\dieterrorprop*

\begin{proof}
Recall the \gls{diet} $p$-value introduced in \cref{alg:diet}:
\begin{align*}
\pval = \frac{1}{M + 1} \left( 1 + \sum_{m=1}^M \mathbbm{1}\left[ \rho(\Dde) \geq \rho(\Dde^{(m)}) \right] \right).
\end{align*}

We will prove that if a $\qhat$ estimator is trained on data $\Dxtyz$, the above $p$-value will be super-uniform.
Using the technique from \citet{candes2018panning}, it suffices to show that the following sequence is exchangeable under the null, conditional on samples of $(\rvz, \rvy)$:
\begin{align*}
\rho(\Dde), \rho(\Dde^{(1)}), \dots, \rho(\Dde^{(M)}).
\end{align*}

Note that $\Dde$, and $\{ (\Dde^{(m)}) \}_{m=1}^M$ are datasets of information residuals.
As such, the above sequence can be rewritten as:
\begin{align*}
\rho(\{ \dhati, \ehati \}_{i=1}^N),
\rho(\{ \dhat^{(i,1)}, \ehat^{(i,1)} \}_{i=1}^N), 
\dots,
\rho(\{ \dhat^{(i,M)}, \ehat^{(i,M)} \}_{i=1}^N)
\end{align*}
where $(\dhat^{(i)}, \ehat^{(i)})$ is the $i$th sample of $\Dde$ and 
$(\dhat^{(i,m)}, \ehat^{(i,m)})$ is the $i$th sample of dataset $\Dde^{(m)}$.
As $\rho$ is deterministic, it suffices to show that the following sequence is exchangeable conditional on $\{ (\rvyi, \rvzi) \}_{i=1}^N$:
\begin{align*}
\{ \dhati, \ehati \}_{i=1}^N,
\{ \dhat^{(i,1)}, \ehat^{(i,1)} \}_{i=1}^N, 
\dots,
\{ \dhat^{(i,M)}, \ehat^{(i,M)} \}_{i=1}^N
\end{align*}

Note that $\dhati, \ehati \sim \qhat(\ehat, \dhat \mid \rvxi, \rvyi, \rvzi)$.
This means that the estimated information residuals can be written as $\dhati, \ehati = h(\balphai, \rvxi, \rvyi, \rvzi; \theta_{\Dxtyz})$, where $h$ is a deterministic function (see appendix A of \citet{trivedi2007copula}), and $\theta_{\Dxtyz}$ denotes the fact that $\hat{q}$ is trained on the dataset $\Dxtyz$.
Rewriting the sampling process as a function of independent noise is similar in spirit to the reparameterization trick used in variational inference \citep{kingma2015variational}.

In this alternative representation, $\balphai$ is a sample of exogenous variable $\balpha$ that represents the noise in $\qhat$.
Using the same notation, $\dhat^{(i,m)}, \ehat^{(i,m)} = h(\balpha^{(i,m)}, \rvxt^{(i,m)}, \rvyi, \rvzi; \theta_{\Dxtyz})$, where $\rvxt^{(i,m)}$ is the $i$th sample of the $m$th null dataset $\rmXt^{(m)}$ and $\balpha^{(i,m)}$ is another independent sample of $\balpha$.
This means the above sequence can be written as:
\begin{align*}
&\{ h(\balphai, \rvxi, \rvyi, \rvzi; \theta_{\Dxtyz}) \}_{i=1}^N,\\
&\{ h(\balphaim{1}, \rvxt^{(i,1)}, \rvyi, \rvzi; \theta_{\Dxtyz}) \}_{i=1}^N, \\
&\vdots \\
&\{ h(\balphaim{M}, \rvxt^{(i,M)}, \rvyi, \rvzi; \theta_{\Dxtyz}) \}_{i=1}^N.
\end{align*}
Since $h$ is deterministic and learned from $\Dxtyz$,
exchangeability of the set of random variables above reduces to exchangeability of the following:
\begin{align}
\label{eq:ex-seq}
\begin{split}
& (\{\balphai, \rvxi, \rvyi, \rvzi\}_{i=1}^N,{\Dxtyz}), \\ 
& (\{ \balphaim{1}, \rvxt^{(i,1)}, \rvyi, \rvzi \}_{i=1}^N, {\Dxtyz}),\\
 \vdots \\
& (\{\balphaim{M}, \rvxt^{(i,M)}, \rvyi, \rvzi \}_{i=1}^N, {\Dxtyz})\end{split}
\end{align}

Now, recall that $\Dxtyz = \{ \rvxti, \rvyi, \rvzi) \}_{i=1}^N$ where each $ \rvxti$ is a random sample from $p(\rvx \g \rvz = \rvzi)$ and note that the only dependence between $\Dxtyz$ and $\rvxi, \rvx^{(i,m)}$ is through $\rvyi,\rvzi$.
Then, collecting the identically distributed samples within in each element in the sequence in \cref{eq:ex-seq}
\[ \rvxi \CI \rvxti \g \rvy_i, \rvz_i \implies \{\rvxi\}_{i=1}^N \CI \Dxtyz \g  \{ (\rvyi, \rvzi) \}_{i=1}^N.\]
\[ \forall m \quad \rvxt^{(i,m)} \CI \rvxti \g \rvy_i, \rvz_i \implies \{\rvxt^{(i,m)}\}_{i=1}^N \CI \Dxtyz \g  \{ (\rvyi, \rvzi) \}_{i=1}^N.\]
This fact imply these two equalities in distribution (between the conditioning set containing $\Dxtyz$ and otherwise):
\begin{align}
\begin{split}
	\label{eq:remove-d}
	\{\rvxi\}_{i=1}^N \g  \{ (\rvyi, \rvzi) \}_{i=1}^N,  \Dxtyz  \quad &=^d \quad \{\rvxi\}_{i=1}^N \g  \{ (\rvyi, \rvzi) \}_{i=1}^N, \\
\forall m \quad \{\rvx^{(i,m)} \}_{i=1}^N \g  \{ (\rvyi, \rvzi) \}_{i=1}^N,  \Dxtyz  \quad &=^d \quad  \{\rvx^{(i,m)} \}_{i=1}^N \g  \{ (\rvyi, \rvzi) \}_{i=1}^N
\end{split}	
\end{align}

Now, the two RHS's above are equal in distribution under the null hypothesis: under $\mathcal{H}_0$, $p(\rvx \mid \rvz) = p(\rvx \mid \rvz, \rvy)$, which means that the distribution of $\rvxi$ is equal to the distribution of $\rvxt^{(i,m)}$ given $\{ (\rvyi, \rvzi) \}_{i=1}^N$.
This fact means the LHS's in \cref{eq:remove-d} are equal which implies the following equality in distribution
\begin{align}
	\label{eq:cond-dataset}
	\{\rvxi\}_{i=1}^N \g  \{ (\rvyi, \rvzi) \}_{i=1}^N,  \Dxtyz  \quad &=^d \quad \{\rvx^{(i,m)} \}_{i=1}^N \g  \{ (\rvyi, \rvzi) \}_{i=1}^N,  \Dxtyz 
\end{align}

Then, recalling that $\balphai$ and $\balphaim{m}$ for all $m$ are exogenous random variables, 
\cref{eq:cond-dataset} implies that  given $\{ (\rvyi, \rvzi) \}_{i=1}^N, \Dxtyz$, the random variable $\{(\balphai, \rvxi)\}_{i=1}^N$ is distributed identically to $\{\balphaim{m},\rvxt^{(i,m)}\}_{i=1}^N$ for any $m$.
Finally, as $\{ (\rvyi, \rvzi) \}_{i=1}^N, \Dxtyz$ is constant across each element of the sequence in \cref{eq:ex-seq}, the sequence is exchangeable.
\end{proof}

\begin{corollary}
    Let $(\rvx, \rvy, \rvz)$ be drawn from any distribution $p(\rvx, \rvy, \rvz)$ and $\Dxyz$
    consist of $N$ iid samples from this distribution.
    If $\rvx \CI \rvy \mid \rvz$, then for any measure of marginal dependence $\rho: (\sR \times \sR)^N \to \sR$, let $\pval$ the $p$-value computed using any residuals $\ehat=u(\rvx, \rvz)$ and $\dhat=v(\rvy, \rvz)$ where $u,v$ are learned from the null-dataset $\Dxtyz\sim p(\rvx\g \rvz)p(\rvz, \rvy)$.
    Then, $\pval$ will stochastically dominate a $\textrm{Uniform}(0,1)$ random variable.
    \label{corr:error-control}
\end{corollary}
\begin{proof}
The one property of \gls{diet} used in proving \cref{prop:type-1-error} is that $\hat{q}$ is learned using the dataset $\Dxtyz$.
This gives us the property that under the null hypothesis
\[\{\rvxi\}_{i}^N \CI \Dxtyz \g \{\rvyi, \rvzi\}_{i}^N \} \qquad \ 
\{\rvxt^{(i,m)}\}_{i}^N \CI \Dxtyz \g \{\rvyi, \rvzi\}_{i}^N \},\]
because 1) the dependence between variables $\rvxi$ or $\rvxt^{(i,m)}$ and  $\Dxtyz=\{\rvxti, \rvyi, \rvzi\}_{i=1}^N$ and is only due to $\{\rvyi, \rvzi\}_{i}^N$, which we condition on, 2) the $\rvxti$ samples in $\Dxtyz$ are independent samples drawn from $p(\rvx \g \rvz=\rvzi)$.
These two properties imply the following independencies:
\[ \rvxi \CI \rvxti \g \rvy_i, \rvz_i \implies \{\rvxi\}_{i=1}^N \CI \Dxtyz \g  \{ (\rvyi, \rvzi) \}_{i=1}^N.\]
\[ \forall m \quad \rvx^{(i,m)} \CI \rvxti \g \rvy_i, \rvz_i \implies \{\rvxt^{(i,m)}\}_{i=1}^N \CI \Dxtyz \g  \{ (\rvyi, \rvzi) \}_{i=1}^N.\]

For any $u,v$ learned from $\Dxtyz\sim p(\rvx\g \rvz)p(\rvz, \rvy)$, the same properties hold because samples from $\hat{q}(\ehat, \dhat \g \rvy, \rvx, \rvz)$ are produced as $\ehati=u(\rvxi, \rvzi)$ and $\dhati=v(\rvyi, \rvzi)$ and $u,v$ are learned from $\Dxtyz$.

Thus, \cref{thm:diet-cdf-power} will hold for any functions $u,v$ learned from $\Dxtyz$ or using data from $p(\rvx, \rvz)$ and $p(\rvy, \rvz)$ respectively; then, using $\ehat=u(\rvx, \rvz)$ and $\dhat=v(\rvy, \rvz)$ to compute a $p$-value using 
\[\pval = \frac{1}{M + 1} \left( 1 + \sum_{m=1}^M \mathbbm{1}\left[ \rho(\Dde) \geq \rho(\Dde^{(m)}) \right] \right)\]
will result in a super-uniform $\pval$.
\end{proof}

\subsection{Proof of \cref{thm:diet-cdf-power}}
\label{sec:proof-of-diet-cdf-power}

\dietpowertheorem*

\begin{proof}

To test the conditional independence relationship $\rvx \CI \rvy \mid \rvz$, \gls{diet} tests the marginal independence between $\beps$ and $\bdelta$.
The aim of this proof is to show that $\beps \CI \bdelta$ if and only if $\rvx \CI \rvy \mid \rvz$.
If this reduction holds, then under the alternate hypothesis $\gH_1$ where $\rvx \not \CI \rvy \mid \rvz$, the distribution of the test statistic $T(\Dxyz)$ will be different from the distribution of each of the null statistics $T(\Dxtyz^{(m)})$.
Then, given any marginal independence test that achieves power $> \alpha$ with statistic $\rho$, \gls{diet} with the same statistic is a \textit{conditional} independence test with power $> \alpha$.

The proof is structured in the following manner.
First, we will show that using the null data $\Dxtyz^{(m)}$, the sampled values of $\beps$ and $\bdelta$ will be independent.
Then, we will show that using the true data $\Dxyz$, the sampled values of $\beps$ and $\bdelta$ will be dependent.
Finally, we discuss how the existence of a marginal independence test with power $> \alpha$ implies that \gls{diet} will also achieve power $> \alpha$ using data $\Dxyz$.

\paragraph{Prerequisites.}

We first outline some properties will be used in both the null statistics and the test statistic section.

\begin{align}
	p(\beps, \rvz) &= \int p(\beps, \bdelta, \rvz) d\bdelta && \text{by marginalization} \nonumber \\
	&= \int p(\beps, \bdelta) p(\rvz) d\bdelta && \text{by data distribution} \nonumber \\
	&= p(\beps) p(\rvz) \nonumber
\end{align}
Thus,
\begin{align}
\beps &\CI \rvz \label{eqn:ehat-indep-z}\\
\bdelta &\CI \rvz \label{eqn:dhat-indep-z}.
\end{align}

\paragraph{Null statistics $T(\Dxtyz^{(m)})$.}

Recall that in each of the null datasets, the following factorization of the data distribution $p(\rvx, \rvy, \rvz)$ holds by construction:
\begin{align}
	p(\rvx, \rvy, \rvz) = p(\rvx \mid \rvz) p(\rvy \mid \rvz) p(\rvz). \label{eqn:factorization-null-data}
\end{align}
We can use this property to make the following sequence of deductions.
Letting $p(\beps, \bdelta, \rvz)$ be the distribution implied by $(\beps, \bdelta, \rvz)$,
\begin{align*}
p(\beps, \bdelta, \rvz)
	&= \int p(\beps, \bdelta \mid \rvx, \rvy, \rvz) p(\rvx, \rvy, \rvz) d\rvx d\rvy  \\ 
	&= 
	\int p(\beps \mid \rvx, \rvz) p(\bdelta \mid \rvy, \rvz) p(\rvx, \rvy, \rvz) d\rvx d\rvy && \text{$\beps$ and $\bdelta$ are each functions of $\rvz$ and either $\rvx$ or $\rvy$} \\
	&= \int p(\beps \mid \rvx, \rvz) p(\bdelta \mid \rvy, \rvz) p(\rvx \mid \rvz) p(\rvy \mid \rvz) p(\rvz) d\rvx d\rvy && \text{by \cref{eqn:factorization-null-data}} \\
	&= \int p(\beps, \rvx \mid \rvz) p(\bdelta, \rvy \mid \rvz) p(\rvz) d\rvx d\rvy \\
	&= p(\beps \mid \rvz) p(\bdelta \mid \rvz) p(\rvz) \\
p(\beps, \bdelta \mid \rvz) &= p(\beps \mid \rvz) p(\bdelta \mid \rvz).
\end{align*}
The distribution of $(\rvy, \rvz)$ under the null is the same as distribution of $(\rvy, \rvz)$ in the data.
Then since $\bdelta \CI \rvz$ (eq. \ref{eqn:dhat-indep-z}) holds in the data distribution, the independence of $\bdelta$ and $\rvz$ also holds under the null distribution \cref{eqn:factorization-null-data}:
\begin{align*}
    \bdelta &\CI \rvz && \text{ where } (\rvx, \rvy, \rvz) \sim p(\rvx \mid \rvz) p(\rvy \mid \rvz) p(\rvz); \quad \bdelta = F_{\rvy \mid \rvz}(\rvy \mid \rvz).
\end{align*}
Using the same logic, \cref{eqn:ehat-indep-z} implies 
\begin{align*}
    \beps &\CI \rvz && \text{ where } (\rvx, \rvy, \rvz) \sim p(\rvx \mid \rvz) p(\rvy \mid \rvz) p(\rvz); \quad \beps = F_{\rvx \mid \rvz}(\rvx \mid \rvz).
\end{align*}
Using the above facts, 
\begin{align*}
	p(\beps, \bdelta) &= \int p(\beps, \bdelta \mid \rvz) p(\rvz) d\rvz && \text{by marginalization}\\
	&= \int p(\beps \mid \rvz) p(\bdelta \mid \rvz) p(\rvz) d\rvz \\
	&= \int p(\beps) p(\bdelta) p(\rvz) d\rvz \\
	&= p(\beps) p(\bdelta).
\end{align*}

Therefore, when using a null dataset $\Dxtyz^{(m)}$, $\beps \CI \bdelta$.

\paragraph{Test statistic $T(\Dxyz)$.}

Under $\gH_1$, $\rvx \not \CI \rvy \mid \rvz$.
In such cases, the sampled values of $\beps$ and $\bdelta$ using $\Dxyz$ must be dependent.
Specifically, the following sequence of implications must hold:
\begin{align*}
	\rvx \not \CI \rvy \mid \rvz
	\Rightarrow
	\beps \not \CI \bdelta \mid \rvz
	\Rightarrow \beps \not \CI \bdelta.
\end{align*}

The first implication follows because both $F_{\rvx \mid \rvz}(\rvx \mid \rvz)$ and $F_{\rvy \mid \rvz}(\rvy \mid \rvz)$ are invertible for any fixed value of $\rvz$.
Next we prove the second implication.
This is equivalent to:
\begin{equation*}
	\bdelta \CI \beps \Rightarrow \bdelta \CI \beps \mid \rvz.
\end{equation*}

We know that $p(\beps, \bdelta \mid \rvz) = p(\beps, \bdelta)$.
It follows that $\bdelta \CI \beps \Rightarrow \bdelta \CI \beps \mid \rvz$:
\begin{align*}
	p(\beps, \bdelta \mid \rvz) &= p(\beps, \bdelta) \\
	&= p(\beps) p(\bdelta) && \text{since } \bdelta \CI \beps \\
	&= p(\beps \mid \rvz) p(\bdelta \mid \rvz) && \text{by \cref{eqn:ehat-indep-z,eqn:dhat-indep-z}}
\end{align*}

We have thus far established that under $\gH_1$, $\bdelta \not \CI \beps$, but under $\gH_0$, $\bdelta \CI \beps$.
Now, consider $\psi(\Dbde, \alpha): (\sR \times \sR)^N \times [0,1] \to \{0, 1\}$, a marginal independence test that uses statistic $\rho: (\sR \times \sR)^N \to \sR$ and has power greater than level $\alpha$.
This means that there exists a rejection region $R_\alpha = \{ \D \in (\sR \times \sR)^N : \psi(\D, \alpha) = 1 \}$ where
$\sP_{\gH_1}(R_\alpha) \geq \sP_{\gH_0}(R_\alpha)$.
In other words, for a sample size of $N$ and statistic $\rho$ there is sufficient evidence to reject the null hypothesis.

Then, \gls{diet} equipped with $\rho$, $F_{\rvx \mid \rvz}(\rvx \mid \rvz)$, and $F_{\rvy \mid \rvz}(\rvy \mid \rvz)$ is a 
conditional independence test $\zeta(\Dxyz, \alpha): (\sR \times \sR \times \sR^{d_\rvz})^N \times [0,1] \to \{0, 1\}$ with rejection region $S_\alpha = \{ \Dxyz \in (\sR \times \sR \times \sR^p)^N : \zeta(\Dxyz, \alpha) = 1 \}$ such that $\sP_{\gH_1}(S_\alpha) \geq \sP_{\gH_0}(S_\alpha)$.
This follows directly from the previous fact because \gls{diet} uses the marginal dependence  $\beps$ and $\bdelta$ to test the conditional independence between $\rvx$ and $\rvy$ given $\rvz$.

Thus, if there is a marginal test that achieves power greater than $\alpha$, then \gls{diet} under the conditions of \cref{thm:diet-cdf-power} will also achieve power greater than $\alpha$.
	
\end{proof}

\subsubsection{Example family of data generating proceses that satisfy the core assumption in \cref{thm:diet-cdf-power}}\label{appsec:example}
Here, we specify a family of data generating processes that satisfies $ (\beps,\bdelta)\CI \rvz$.
Let $\rve,\rvd$ be any continuously distributed random variables with contiguous support and let the joint distribution over $\rve, \rvd,\rvz $ be 
\[p(\rve,\rvd, \rvz) = p(\rve, \rvd) p (\rvz).\]
For any pair of functions $f(\cdot, \cdot),g(\cdot, \cdot)$ that are continuous and strictly monotonic in the first argument, let samples from $p(\rvx, \rvy, \rvz) $ be generated as:
\[\rvz \sim p(\rvz), \,\, \rve,\rvd \sim p(\rve,\rvd), \,\, \rvx = f(\rve, \rvz), \,\, \rvy = g(\rvd, \rvz).\]

The core assumption holds on all $p(\rvx, \rvy, \rvz) $ with the above form.
First, we express $\beps$ as a deterministic function of $\rve$: almost surely under $p(\rvy, \rvx, \rvz)$: with $f_z^{-1}(x, z)$ as the inverse of the function $f$ in the first argument with $z$ fixed
\[ \beps = F_{\rvx\mid \rvz}(\rvx \mid \rvz) = F_{f_z^{-1}(\rvx, \rvz)\mid \rvz}(f_z^{-1}(\rvx, \rvz) \mid \rvz) = F_{\rve \mid \rvz}(\rve \mid \rvz) = F_{\rve}(\rve).\]

The second equality holds as \acrshortpl{cdf} are invariant to strictly monotonic transformations of the underlying random variables and the fourth due to the independence $\rve \CI \rvz$.
Similarly $\bdelta=F_\rvd(\rvd)$.
In turn, $\bdelta,\beps$ are deterministic functions of $\rvd, \rve$ respectively, and the core assumption holds:
\[(\rvd, \rve)\CI \rvz \implies (F_{\rve}(\rve), F_{\rvd}(\rvd)) \CI \rvz \implies (\beps, \bdelta) \CI \rvz.\]

\subsection{Generalizing assumptions for distillation}
\label{sec:proof-of-sufficient-conditions-for-power}

In this section we consider data generating processes of the following form:
\begin{align*}
	\rvz \sim p(\rvz) \qquad (\rve, \rvd) \sim p(\rve, \rvd) \qquad \rvx = f(\rve, \rvz) \qquad \rvy = g(\rvd, \rvz).
\end{align*}
The goal of a distillation procedure like \gls{diet} or the $d_0$-\gls{crt} is to first estimate $\rve$ and $\rvd$ from samples of $(\rvx, \rvy, \rvz)$, then test the marginal independence of these estimates: $\beps \CI \bdelta$.
Since $\rve$ and $\rvd$ are unobserved, samples in $\Dxyz$ map to a distribution $p(\rvd, \rve \mid \rvx, \rvy, \rvz)$ over the possible values of $(\rve, \rvd)$.
The distribution $p(\rve, \rvd \mid \rvx, \rvy, \rvz)$ is also unknown; it must be estimated using an estimator $\qhat(\beps, \bdelta \mid \rvx, \rvy, \rvz)$.

However, not all $\beps, \bdelta \sim \qhat(\beps, \bdelta \mid \rvx, \rvy, \rvz)$ will yield power to reject the null hypothesis $\gH_0: \rvx \CI \rvy \mid \rvz$.
In some cases $\beps \CI \bdelta$ but $\rvx \not \CI \rvy \mid \rvz$.
Consider this example from \citet{puli2020general}.
Let $\rvx = \beps$ and $\rvy = \bdelta$, let $\beps \sim \text{Uniform}(0, 1)$, $\bdelta \sim \text{Uniform}(0, 1)$, and
\begin{align*}
	\rvz = \begin{cases}
		\beps + \bdelta & \text{if } \beps + \bdelta \leq 1 \\
		\beps + \bdelta - 1 & \text{otherwise} 
	\end{cases}.
\end{align*}
In this example, $\beps$ and $\bdelta$ are independent of each other, but $\rvx$ and $\rvy$ are clearly dependent given $\rvz$.
The following theorem, \cref{thm:diet-general}, gives sufficient conditions on $\qhat(\beps, \bdelta \mid \rvx, \rvy, \rvz)$ to ensure that $\beps \CI \bdelta$ if and only if $\rvx \CI \rvy \mid \rvz$.
We later show that the only way to satisfy the conditions in \cref{thm:diet-general} are through assumptions on the data generating process.

\setcounter{theorem}{1}

\begin{theorem}
Consider a data generating process of the following form:
\[\rvz \sim p(\rvz), \quad \rve, \rvd \sim p(\rve, \rvd), \quad \rvx = f(\rve,\rvz) \quad \rvy = g(\rvd,\rvz).\]
Let $\beps, \bdelta$ be distributed according to:
\[(\beps, \bdelta, \rvx, \rvy, \rvz) \sim \qhat(\beps, \bdelta \mid \rvx, \rvy, \rvz) p(\rvx, \rvy, \rvz).
\]
Further let,
\begin{align}
	\qhat&(\beps, \bdelta \mid \rvx, \rvy, \rvz) = p(\beps \mid \rvx, \rvz) p(\bdelta \mid \rvy, \rvz), \tag{factorization} \label{eqn:factorization} \\
	\exists &\tilde{f}, \tilde{g} \quad 
	\quad \rvx \overset{\text{a.s.}}{=} \tilde{f}(\beps, \rvz), \quad \rvy \overset{\text{a.s.}}{=} \tilde{g}(\bdelta, \rvz), \tag{reconstruction} \label{eqn:reconstruction} \\
	(&\rvd, \bdelta) \CI \rvz \qquad (\rve, \beps) \CI \rvz.
	 \tag{joint independence} \label{eqn:joint-independence}
\end{align}
Let $\psi(\Dbde, \alpha): (\sR \times \sR)^N \times [0,1] \to \{0, 1\}$ be a marginal independence test that uses statistic $\rho: (\sR \times \sR)^N \to \sR$ and has power greater than $\alpha$.
Let $\Dbde$ be a dataset of $N$ samples of $(\beps, \bdelta)$ generated using $\qhat(\beps, \bdelta \mid \rvx, \rvy, \rvz)$ and $\Dxyz$.
Then, $\psi$ using $\Dbde$ and $\rho$ is also a conditional test of independence for $\rvx \CI \rvy \mid \rvz$ with power greater than $\alpha$.
\label{thm:diet-general}
\end{theorem}

\begin{proof}

The core of this proof is to show that if \ref{eqn:factorization}, \ref{eqn:reconstruction}, and \ref{eqn:joint-independence} are satisfied,
then
\begin{equation*}
	\rvx \CI \rvy \mid \rvz \Leftrightarrow \beps \CI \bdelta.
\end{equation*}
If this reduction is possible, then under $\gH_1$, $\beps \not \CI \bdelta$, but under $\gH_0$, $\beps \CI \bdelta$.
This implies that the distribution of the marginal dependence test statistic $\rho(\Dde)$ is different from that of each null statistic $\rho(\Dde^{(m)})$.
Thus, the $p$-value computed by $\psi$ will be close to 0:
\begin{align*}
	\pval = \frac{1}{M+1}\left(1 + \sum_{m=1}^M \mathbbm{1}(\rho(\Dde) \leq \rho(\Dde^{(m)})) \right).
\end{align*}

Let $p(\rvd, \rve, \beps, \bdelta, \rvz)$ be a distribution over variables $\rvd, \rve, \beps, \bdelta, \rvz$.
The variables $\bdelta$ and $\beps$ are samples from $\qhat(\beps, \bdelta \mid \rvx, \rvy, \rvz)$.
For simplicity, we show the proof of \cref{thm:diet-general} when all random variables are continuous, but the same reasoning holds for discrete random variables.

\paragraph{Null statistics.}

For null statistics $\rho(\Dde^{(m)})$ computed using null data $\Dxtyz^{(m)}$, $\bdelta, \beps \sim \qhat(\bdelta, \beps \mid \rvxt, \rvy, \rvz)$ must be independent.
In the null data, $\rvxt \CI \rvy \mid \rvz$ by construction, so
the following must hold:
\begin{equation}
	\rvxt \CI \rvy \mid \rvz
	\Rightarrow \beps \CI \bdelta.
	\label{eqn:h0-implications}
\end{equation}
We show this fact by manipulating the distribution $\qhat(\beps, \bdelta \mid \rvxt, \rvy, \rvz) p(\rvxt, \rvy, \rvz)$. 
In this proof, we write $\qhat(\beps, \bdelta \mid \rvxt, \rvy, \rvz)$ as $p(\beps, \bdelta \mid \rvxt, \rvy, \rvz)$ to simplify the notation:
\begin{align*}
p(\beps, \bdelta, \rvz)
	&= \int p(\beps, \bdelta \mid \rvxt, \rvy, \rvz) p(\rvxt, \rvy, \rvz) d\rvxt d\rvy  \\ 
	&= 
	\int p(\beps \mid \rvxt, \rvz) p(\bdelta \mid \rvy, \rvz) p(\rvxt, \rvy, \rvz) d\rvxt d\rvy && \text{By \ref{eqn:factorization}} \\
	&= \int p(\beps \mid \rvx, \rvz) p(\bdelta \mid \rvy, \rvz) p(\rvxt \mid \rvz) p(\rvy \mid \rvz) p(\rvz) d\rvxt d\rvy && \text{In $\Dxtyz$ } \rvxt \CI \rvy \mid \rvz \\
	&= \int p(\beps, \rvxt \mid \rvz) p(\bdelta, \rvy \mid \rvz) p(\rvz) d\rvxt d\rvy \\
	&= p(\beps \mid \rvz) p(\bdelta \mid \rvz) p(\rvz)
\end{align*}
Consequently, $p(\beps, \bdelta \mid \rvz) = p(\bdelta \mid \rvz) p(\bdelta \mid \rvz) \Leftrightarrow \beps \CI \bdelta \mid \rvz$.
Here, if $\bdelta \CI \rvz$ and $\beps \CI \rvz$, then
\[\beps \CI \bdelta \mid \rvz \Rightarrow \beps \CI (\bdelta, \rvz) \Rightarrow \beps \CI \bdelta\]
Thus, if $\rvxt \CI \rvy \mid \rvz$, as is the case in the computation of each of the null statistics $\rho(\Dde^{(m)})$, then for $\bdelta, \beps \sim \qhat(\bdelta, \beps \mid \rvxt, \rvy, \rvz)$, $\bdelta \CI \beps$.

\paragraph{Test statistic under $\gH_1$.}

For the test statistic $\rho(\Dde)$ computed using the data $\Dxyz$, $\bdelta$ and $\beps$ must be dependent.
Under $\gH_1$, $\rvx \not \CI \rvy \mid \rvz$, so
the following sequence of implications must hold:
\begin{equation}
	\rvx \not \CI \rvy \mid \rvz
	\Rightarrow
	\tilde{f}(\beps, \rvz) \not \CI \tilde{g}(\bdelta, \rvz) \mid \rvz
	\Rightarrow
	\beps \not \CI \bdelta \mid \rvz
	\Rightarrow \beps \not \CI \bdelta.
	\label{eqn:h1-implications}
\end{equation}

The first implication follows directly from \ref{eqn:reconstruction}, the second holds because $\beps $ and $\bdelta$ are the only sources of variance when $\rvz$ is fixed.
This last implication is equivalent to the following statement, which we will subsequently prove:
\begin{equation*}
	\bdelta \CI \beps \Rightarrow \bdelta \CI \beps \mid \rvz.
\end{equation*}

First, note the following properties.
Using \ref{eqn:joint-independence}, we show that the distribution $p(\bdelta, \rvz)$ factorizes, implying that $\bdelta$ and $\rvz$ are marginally independent:
\begin{gather}
p(\bdelta, \rvz) = 
	\int p(\rvd, \bdelta, \rvz) d\rvd = \int p(\rvd, \bdelta) p(\rvz) d\rvd = p(\bdelta) p(\rvz), \label{eqn:joint-independence-equals-marginal-independence} \\
p(\beps, \rvz) = 
	\int p(\rve, \beps, \rvz) d\rvd = \int p(\rve, \beps) p(\rvz) d\rve = p(\beps) p(\rvz). \label{eqn:joint-independence-equals-marginal-independence-e}
\end{gather}
Further, \ref{eqn:joint-independence} implies the following:
\begin{align}
	\frac{p(\bdelta, \rvd, \rvz)}{p(\rvd)} &= \frac{p(\bdelta, \rvd) p(\rvz)}{p(\rvd)} = p(\bdelta \mid \rvd) p(\rvz) = p(\bdelta \mid \rvd) p(\rvz \mid \rvd) && \text{Since $\rvz \CI \rvd$ by definition} \label{eqn:dhat-z-given-delta-factorization} \\
		\frac{p(\beps, \rve, \rvz)}{p(\rve)} &= \frac{p(\beps, \rve) p(\rvz)}{p(\rve)} = p(\beps \mid \rve) p(\rvz) = p(\beps \mid \rve) p(\rvz \mid \rve) && \text{Since $\rvz \CI \rve$ by definition} \label{eqn:ehat-z-given-epsilon-factorization}
\end{align}

Next, note that \ref{eqn:factorization} implies:
\begin{align}
	p(\beps, \bdelta \mid \rvz, \rve, \rvd) &= p(\beps, \bdelta \mid \rvx, \rvy, \rvz, \rve, \rvd) \nonumber && \text{$\rvx$ and $\rvy$ are fully determined by $(\rvz, \rvd, \rve)$} \\
	&= p(\beps, \bdelta \mid \rvx, \rvy, \rvz) && \text{$(\beps, \bdelta)$ are functions of only $(\rvx, \rvy, \rvz)$ and exogenous noise} \nonumber \\
	 &=  p(\beps \mid \rvx, \rvz) p(\bdelta \mid \rvy, \rvz) && \text{\ref{eqn:factorization} assumption} \nonumber \\
	&=  p(\beps \mid \rvx, \rvz,  \rve) p(\bdelta \mid \rvy, \rvz,  \rvd) && 
	\text{$\beps, \bdelta$ are functions of $(\rvx, \rvz)$ and $(\rvy, \rvz)$ respectively and exogenous noise} \nonumber \\
	&=  p(\beps \mid \rve, \rvz) p(\bdelta \mid \rvd, \rvz) && \text{$\rvx = f(\rve, \rvz)$, $\rvy = g(\rvd, \rvz)$} \label{eqn:factorization-epshat-deltahat}.
\end{align}

Using the above facts, we then show that $p(\beps, \bdelta \mid \rvz) = p(\beps, \bdelta)$:
\begin{align*}
	p(\beps, \bdelta \mid \rvz) &= \int p(\beps, \bdelta, \rve, \rvd \mid \rvz) d\rve d\rvd && \text{By marginalization} \\
	 &= \int p(\beps, \bdelta \mid \rvz, \rve, \rvd) p(\rve , \rvd \mid \rvz) d\rve d\rvd \\
	 &= \int p(\beps, \bdelta \mid \rvz, \rve, \rvd) p(\rve , \rvd) d\rve d\rvd && \text{By definition of the data generating process} \\
	 &= \int p(\beps \mid \rvz, \rve)  p(\bdelta\beps \mid \rvz, \rvd) p(\rve , \rvd) d\rve d\rvd && \text{By \cref{eqn:factorization-epshat-deltahat}} \\
	 &= \int p(\beps \mid \rve)  p(\bdelta \mid \rvd) p(\rve , \rvd) d\rve d\rvd && \text{By \cref{eqn:dhat-z-given-delta-factorization,eqn:ehat-z-given-epsilon-factorization}}
\end{align*}

\begin{align*}
	p(\beps, \bdelta) &= \int p(\beps, \bdelta \mid \rvz, \rve, \rvd) p(\rvz, \rve, \rvd)   d\rvz d\rve d\rvd && \text{By marginalization} \\
	&= \int p(\beps \mid \rvz, \rve)  p(\bdelta \mid \rvz, \rvd) p(\rvz, \rve, \rvd) d\rvz d\rve d\rvd && \text{By \cref{eqn:factorization-epshat-deltahat}} \\
	&= \int p(\beps \mid \rve)  p(\bdelta \mid \rvd) p(\rvz, \rve, \rvd) d\rvz d\rve d\rvd && \text{By \cref{eqn:dhat-z-given-delta-factorization,eqn:ehat-z-given-epsilon-factorization}} \\
	&= \int p(\beps \mid \rve)  p(\bdelta \mid \rvd) p(\rve, \rvd \mid \rvz) p(\rvz) d\rvz d\rve d\rvd \\
	&= \int p(\beps \mid \rve)  p(\bdelta \mid \rvd) p(\rve, \rvd) p(\rvz) d\rvz d\rve d\rvd && \text{By definition of the data generating process} \\
	&= \int p(\beps \mid \rve)  p(\bdelta \mid \rvd) p(\rve, \rvd) \left( \int p(\rvz) d\rvz \right) d\rve d\rvd \\
	&= \int p(\beps \mid \rve)  p(\bdelta \mid \rvd) p(\rve, \rvd) d\rve d\rvd
\end{align*}

Using all of the above facts, it follows that 
$\bdelta \CI \beps \Rightarrow \bdelta \CI \beps \mid \rvz$:
\begin{align*}
	p(\beps, \bdelta \mid \rvz) &= p(\beps, \bdelta) \\
	&= p(\beps) p(\bdelta) && \text{Since } \bdelta \CI \beps \\
	&= p(\beps \mid \rvz) p(\bdelta \mid \rvz) && \text{By \cref{eqn:joint-independence-equals-marginal-independence,eqn:joint-independence-equals-marginal-independence-e}},
\end{align*}
thus satisfying the sequence of implications in \cref{eqn:h1-implications}.

We have thus far established that under $\gH_1$, $\bdelta \not \CI \beps$, but under $\gH_0$, $\bdelta \CI \beps$.
Therefore, given a marginal independence test $\psi(\Dde, \alpha): (\sR \times \sR)^N \times [0,1] \to \{0, 1\}$ that is known to achieve power greater than level $\alpha$, using $\psi$ with a dataset of samples from $\qhat(\beps, \bdelta \mid \rvx, \rvy, \rvz)$ will result in a conditional test with power greater than $\alpha$.

\end{proof}

\subsection{Proof of \cref{prop:ump-info-residuals}}
\label{sec:proof-ump-info-residuals}

\dietumpprop*

\begin{proof}

Using the conditional \glspl{cdf} $F_{\rvx \mid \rvz}(\rvx \mid \rvz)$ and $F_{\rvy \mid \rvz}(\rvy \mid \rvz)$,
define the following terms for convenience:
\begin{align*}
	\fz(\rvx) &:= F_{\rvx \mid \rvz}(\rvx \mid \rvz) \\
	\gz(\rvy) &:= F_{\rvy \mid \rvz}(\rvy \mid \rvz) \\
	J &= 
    \begin{bmatrix} 
\frac{d}{d\rvx} \fz(\rvx) & \frac{d}{d\rvy} \fz(\rvx)\\
\frac{d}{d\rvx} \gz(\rvy) & \frac{d}{d\rvy} \gz(\rvy)
\end{bmatrix} \\
&=
\begin{bmatrix}
\fz'(\rvx) & 0\\
0 & \gz'(\rvy)
\end{bmatrix}.
\end{align*}
The off-diagonals of $J$ are 0 because $\fz(\rvx)$ is not a function of $\rvy$ and $\gz(\rvy)$ is not a function of $\rvx$.
Then using change of variables, we can write:
\begin{align*}
    p(\rvx, \rvy \mid \rvz) &= p(\beps = \fz(\rvx), \bdelta = \gz(\rvy) \mid \rvz) \cdot | \det(J) | \\
    &= p(\beps = \fz(\rvx), \bdelta = \gz(\rvy) \mid \rvz) \cdot | \fz'(\rvx) \cdot \gz'(\rvy) | \\
    &= p(\beps = \fz(\rvx), \bdelta = \gz(\rvy) \mid \rvz) \cdot \fz'(\rvx) \cdot \gz'(\rvy) \\
    &= p(\beps = \fz(\rvx), \bdelta = \gz(\rvy)) \cdot \fz'(\rvx) \cdot \gz'(\rvy).
\end{align*}
The second last step follows because $\fz$ and $\gz$ are monotonically non-decreasing, meaning their derivatives with respect to $\rvx$ or $\rvy$ for a fixed $\rvz$ are non-negative.
The absolute value of the product of two non-negative quantities is just the product of the two quantities.
The last step uses the assumption that $(\beps, \bdelta) \CI \rvz$.
Using similar reasoning,
\begin{align*}
p(\rvx \mid \rvz) &= p(\beps = \fz(\rvx) \mid \rvz) \cdot \fz'(\rvx) = p(\beps = \fz(\rvx)) \cdot \fz'(\rvx) \\
p(\rvy \mid \rvz) &= p(\bdelta = \gz(\rvy) \mid \rvz) \cdot \gz'(\rvy) = p(\bdelta = \gz(\rvy)) \cdot \gz'(\rvy).
\end{align*}

Using the above change of variable results, we can manipulate the likelihood ratio statistic that \citet{katsevich2020theoretical} prove is the conditionally most powerful against point alternatives.

\begin{align*}
    \frac{1}{N} \sum_{i=1}^N \log \frac{
		p(\rvyi \mid \rvxi,  \rvzi)
	}{
		p(\rvyi \mid \rvzi)
	} &= \frac{1}{N} \sum_{i=1}^N \log \frac{
		p(\rvxi, \rvyi \mid \rvzi)
	}{
		p(\rvxi \mid \rvzi) p(\rvyi \mid \rvzi)
	} \\
	&= \frac{1}{N} \sum_{i=1}^N 
	\log \frac{
	    p(\beps = \fzi(\rvxi), \bdelta = \gzi(\rvyi)) \cdot \fzi'(\rvxi) \cdot \gzi'(\rvyi)
	    }{
	    p(\beps = \fz(\rvx)) \cdot \fz'(\rvx) \cdot p(\bdelta = \gz(\rvy)) \cdot \gz'(\rvy)
	    }  \\
	&= \frac{1}{N} \sum_{i=1}^N 
	\log \frac{
	p(\beps = \fzi(\rvxi), \bdelta = \gzi(\rvyi))
	}{
	p(\beps = \fz(\rvx)) \cdot p(\bdelta = \gz(\rvy))
	}.
\end{align*}

Note that this final term on is exactly the mutual-information based marginal dependence measure in the statement of \cref{prop:ump-info-residuals}.
Therefore, the optimal \acrshort{diet} solution is the most powerful conditionally valid test against point alternatives.

\end{proof}

\subsection{\Glspl{dcrt} satisfies conditions in \cref{thm:diet-general} for additive data generating processes}
\label{sec:dcrt-fits-into-our-theorem}

Recall the conditions on $\qhat(\beps, \bdelta \mid \rvx, \rvy, \rvz)$ in \cref{thm:diet-general} required for a conditional independence test to have power greater than its size:

\begin{gather}
	\qhat(\beps, \bdelta \mid \rvx, \rvy, \rvz) = p(\beps \mid \rvx, \rvz) p(\bdelta \mid \rvy, \rvz), \tag{factorization} \\
	\exists \tilde{f}, \tilde{g} \quad \text{s.t.} 
	\quad \rvx \overset{\text{a.s.}}{=} \tilde{f}(\beps, \rvz), \quad \text{and} \quad \rvy \overset{\text{a.s.}}{=} \tilde{g}(\bdelta, \rvz), \tag{reconstruction} \\
	(\rvd, \bdelta) \CI \rvz \qquad (\rve, \beps) \CI \rvz.
	  \tag{joint independence}
\end{gather}

We will show that the $d_0$-\gls{crt}, which reduces a conditional test of independence to a marginal one, satisfies the conditions in \cref{thm:diet-general} under additive noise assumptions on the data generating process.
The $d_I$-\gls{crt} is not discussed here, as it does not reduce $\rvx \CI \rvy \mid \rvz$ to marginal test of independence between two univariate quantities $\beps \CI \bdelta$.

Recall that the $d_0$-\gls{crt} computes the marginal independence between:
\begin{align*}
	\beps = \rvx - \E[\rvx \mid \rvz] \text{ and } \bdelta = \rvy - \E[\rvy \mid \rvz].
\end{align*}
Since $\beps$ only depends on $(\rvx, \rvz)$ and  $\bdelta$ only depends on $(\rvy, \rvz)$, \ref{eqn:factorization} is satisfied.
Next, consider an additive generating process
\begin{align*}
	\rvx &= \bar{f}(\rvz) + \rve \\
	\rvy &= \bar{g}(\rvz) + \rvd,
\end{align*}
where $\rve, \rvd \CI \rvz$
Without loss of generality, $\rve$ and $\rvd$ can have zero expectation.
Under the assumption of additivity, $\rvx - \E[\rvx \mid \rvz] = \rvx - \bar{f}(\rvz) = \rve$ and $\rvy - \E[\rvy \mid \rvz] = \rvy - \bar{g}(\rvz) = \rvd$.
Therefore, computing $\beps$ and $\bdelta$ recovers $\rve$ and $\rvd$ exactly.

Recovering $\rve$ and $\rvd$ exactly also implies that \ref{eqn:reconstruction} and \ref{eqn:joint-independence} are satisfied.
This is because with knowledge of $\rvz$ and $\rve$, $\rvx$ can be reconstructed exactly.
The same holds for $\rvy$ with $\rvz$ and $\rvd$.
The \ref{eqn:joint-independence} property holds trivially because $\beps = \rve$ and $\bdelta = \rvd$, and $\rve$ and $\rvd$ are independent of $\rvz$ by definition.

Thus, all three conditions in \cref{thm:diet-general} are satisfied by the $d_0$-\gls{crt} under additivity.

%% file: sections/appendix_experimental_details.tex
\label{sec:experimental-details}

All experiments are run using a single Intel Xeon Platinum 8268 2.9GHz CPU and an NVIDIA RTX8000 GPU.

\subsection{Training and hyperparameter details for \gls{diet} with \glspl{mdn}}
\label{sec:diet-mdn-details}

To fit the \glspl{mdn}, we use the following network architecture to model each of $\Qcdf{{\rvy \mid \rvz}}(\rvy \mid \rvz ; \theta)$ and $\Qcdf{{\rvx \mid \rvz}}(\rvx \mid \rvz ; \eta)$.
We give details about modelling $\Qcdf{{\rvy \mid \rvz}}(\rvy \mid \rvz ; \theta)$, but the model for $\Qcdf{{\rvx \mid \rvz}}(\rvx \mid \rvz ; \eta)$ is identical.
The network consists of six consecutive fully-connected layers each followed by batch normalization and ReLU activation.
For each input $\rvzi$, the neural network outputs mixture parameters $\pi_\theta$, mean parameters $\mu_\theta$, and variance parameters $\sigma_\theta$, each consisting of $K$ dimensions.
Then, the log-likelihood of $\rvyi \mid \rvzi$ is computed as:
\begin{align*}
    \log \sum_{k=1}^K \pi_\theta^{(k)} \mathcal{N}(\rvyi ; \mu^{(k)}_\theta, \sigma^{(k)}_\theta).
\end{align*}
The training objective involves maximizing average of this quantity over all samples $(\rvzi, \rvyi)$ in the dataset with respect to the parameters $\theta := \{\pi_\theta, \mu_\theta, \sigma_\theta \}$.
This is shown in \cref{eqn:dhat-mle}.
Letting $\Phi$ be the \gls{cdf} of a standard normal random variable, the empirical \gls{cdf} implied by parameters $\theta$ evaluated at a point $(\rvzi, \rvyi)$ is:
\begin{align*}
    \Qcdf{{\rvy \mid \rvz}}(\rvy = \rvyi \mid \rvz = \rvzi ; \theta) = \sum_{k=1}^K \pi_\theta^{(k)} \Phi \left(\frac{\rvyi - \mu^{(k)}_\theta}{\sigma^{(k)}_\theta} \right).
\end{align*}
We employ the Adam \citep{kingma2014adam} optimizer with an initial learning rate of $1 \times 10^{-3}$.
In our experiments, we fix $K = 10$.
Our choice of marginal dependence statistic $\rho$ discretizes $\ehat$ and $\dhat$, then applies the adjusted mutual information estimator from \citet{vinh2009information}.

\subsection{Training and hyperparameter details for baseline \glspl{crt}}
\label{sec:baseline-details}

\paragraph{Test statistic for $d_0$-\gls{crt}.}

In this section, we review the full $p$-value computation for $d_0$-\glspl{crt}.
We implement the Lasso-based models prescribed by \citet{liu2020fast}.
This involves first fitting two regressions with $\ell_1$ regularization:
\begin{align*}
    \argmin_\theta \sum_{i=1}^N (\rvyi - \rvzi \theta )^2 + \lambda_\theta || \theta ||_1, \quad
    \argmin_\eta \sum_{i=1}^N (\rvxti - \rvzi \eta )^2 + \lambda_\eta || \eta ||_1.
\end{align*}
The regularization coefficients $\lambda_\theta$ and $\lambda_\eta$ are found using 5-fold cross-validation.
The test statistics $T(\Dxyz)$ and $T(\Dxtyz^{(m)})$ are computed as follows:
\begin{align*}
    T(\Dxyz) &= \left( \frac{\sum_{i=1}^N (\rvyi - \rvzi \theta)(\rvxi - \rvzi \eta)}{\sum_{i=1}^N (\rvxi - \rvzi \eta)^2} \right)^2 \\
    T(\Dxtyz^{(m)}) &= \left( \frac{\sum_{i=1}^N (\rvyi - \rvzi \theta)(\rvxt^{(i,m)} - \rvzi \eta)}{\sum_{i=1}^N (\rvxt^{(i,m)} - \rvzi \eta)^2} \right)^2,
\end{align*}
where $\rvxt^{(i,m)}$ is the $i$th sample of $\rvxt$ in $\Dxtyz$.
Finally, the $p$-value for the $d_0$-\gls{crt} is computed as:
\begin{align*}
    \frac{1}{M + 1} \left(1 + \sum_{m=1}^M \mathbbm{1}(T(\Dxyz) \leq T(\Dxtyz^{(m)})) \right).
\end{align*}

\paragraph{Test statistic for $d_I$-\gls{crt}.}
In this section, we review the full $p$-value computation for $d_I$-\glspl{crt}.
We implement the method used in \citet{liu2020fast}.
First, the following regressions are fit:
\begin{align*}
    \argmin_\theta \sum_{i=1}^N (\rvyi - \rvzi \theta )^2 + \lambda_\theta || \theta ||_1, \quad
    \argmin_\eta \sum_{i=1}^N (\rvxti - \rvzi \eta )^2 + \lambda_\eta || \eta ||_1.
\end{align*}
The regularization coefficients $\lambda_\theta$ and $\lambda_\eta$ are found using 5-fold cross-validation.

The test statistic $T(\Dxyz)$ is computed in the following manner.
First, the ``top $k$'' dimensions in $\rvz$ are selected using a Lasso heuristic.
Let the set of the top $k$ dimensions be called $S_k$.
The dimensions of $\rvz$ in $S_k$ are those with the highest corresponding $|\theta_j|$, where $\theta_j$ is the $j$th coordinate of $\theta$.
The $d_I$-\gls{crt} then fits a model from $(\rvx - d_\rvx, d_\rvy, \rvz_{\mathrm{top}(k)})$ to $\rvy$.
To explicitly involve first-order interactions, the $d_I$-\gls{crt} we implement includes interaction terms between $(\rvx - d_\rvx)$ and each $\rvz_j \in \rvz_{\mathrm{top}(k)}$.
Using these interaction terms, the following regression is fit:
\begin{align*}
    \argmin_{\beta, \{ \beta_j \}_{j \in S_k} } \sum_{i=1}^N \left( (\rvyi - \rvzi \theta) - \beta (\rvxi - \rvzi \eta) - \sum_{j \in S_k} \beta_j \rvzi_j (\rvxi - \rvzi \eta) \right)^2 + \lambda(|\beta| + \sum_j|\beta_j|).
\end{align*}
The $\ell_1$ penalty coefficient $\lambda$ is chosen through cross validation from amongst $\{10^{-3},10^{-2},10^{-1},1,10^{1},10^{-3},10^{2},10^{3}\}$.
Finally, $T(\Dxyz) := \beta^2 + \frac{1}{k}\sum_{j \in S_k} \beta_j^2$.
This second regression is fit during each evaluation of the test statistic on dataset $\Dxyz$.

The test statistic $T(\Dxtyz^{(m)})$ is computed identically, but with samples from $\Dxtyz^{(m)}$ instead.
The $p$-value is computed in the same way as the $d_0$-\gls{crt}.
Since the Lasso heuristic requires a choice of hyperparameter $k$, we use $k = 2 \log d_\rvz$, where $d_\rvz$ is the number of coordinates in $\rvz$, as recommended by \citet{liu2020fast}.

\paragraph{Test statistic for \gls{hrt}.}

In this section, we review the full $p$-value computation for the \glspl{hrt} used in our experiments.
We use the cross-validated \gls{hrt} from \citet{tansey2018holdout}, who show it achieves higher power than the standard \gls{hrt}.
First, the dataset $\Dxyz$ is split in half into a train and test set: $\Dxyz^{(\textrm{train})}$ and $\Dxyz^{(\textrm{test})}$.
The null datasets $\{ \Dxtyz^{(m, \textrm{train})} \}_{m=1}^M$ are correspondingly split into sets $\{ \Dxtyz^{(m, \textrm{train})} \}_{m=1}^M$ and $\{ \Dxtyz^{(m, \textrm{test})} \}_{m=1}^M$.
Then, the model $\qmodel(\rvy \mid \rvx, \rvz)$, a neural network in this case, is fit using $\Dxyz^{(\textrm{train})}$.
We use the same training setup as with the \glspl{mdn} in \gls{diet}.
$P$-values are then computed using only the test sets.

To compute $T(\Dxyz^{(\textrm{test})})$, we let:
\begin{align*}
    T(\Dxyz^{(\textrm{test})}) = 
    \frac{1}{N/2} \sum_{i=1}^N \mathcal{L}(\qmodel, \rvyi_{\text{test}}, \rvxi_{\text{test}}, \rvzi_{\text{test}}),
\end{align*}
where $\mathcal{L}$ is a loss function evaluated using $\qmodel$ and a sample from $\Dxyz^{(\textrm{test})}$.
When response $\rvy$ is a continuous random variable:
\begin{align*}
    \mathcal{L}(\qmodel, \rvyi_{\text{test}}, \rvxi_{\text{test}}, \rvzi_{\text{test}}) = (\rvyi - \hat{\rvy}^{(i)})^2,
\end{align*}
where $\hat{\rvy}^{(i)}$ is the predicted value of $\qmodel(\rvy \mid \rvx = \rvxi_{\text{test}}, \rvz = \rvzi_{\text{test}})$.
If $\rvy$ is discrete, the loss function is the log-probability of observing $\rvy$ given $\rvx$ and $\rvz$:
\begin{align*}
    \mathcal{L}(\qmodel, \rvyi_{\text{test}}, \rvxi_{\text{test}}, \rvzi_{\text{test}}) = \log \qmodel(\rvy = \rvyi_{\text{test}} \mid \rvx = \rvxi_{\text{test}}, \rvz = \rvzi_{\text{test}}).
\end{align*}
The null statistic $T(\Dxtyz^{(\textrm{test})})$ is computed in a similar way with the same $\qmodel$.

Next, a $p$-value, $\pval_1$, of the \gls{hrt} is computed by
\begin{align*}
    \frac{1}{M + 1} \left(1 + \sum_{m=1}^M \mathbbm{1}(T(\Dxyz^{(\textrm{test})}) \geq T(\Dxtyz^{(m, \textrm{test})})) \right).
\end{align*}
Finally, to compute a cross-validated $p$-value using the \gls{hrt}, we repeat all the steps above to obtain another $p$-value $\pval_2$, but exchanging the roles of the train and test sets.
These two $p$-values $\pval_1$ and $\pval_2$ are combined by taking $\min(1, 2 \cdot \min(\pval_1, \pval_2))$.

\subsection{Variable selection experimental details}
\label{sec:cvs-details}

In this section, we provide specific implementation details for our variable selection experiments.
First recall the setup for these experiments.
Given a set of $d$ covariates $\rvx = \{\rvx_1, \dots, \rvx_d \}$ and a response $\rvy$, we test the conditional independence of each coordinate  $\rvx_j$ with $\rvy$ having observed all other coordinates of $\rvx_{-j}$.
For simplicity, we focus on the \gls{ci} test for only a single coordinate $\rvx_j$ in this section.
The procedure for the other coordinates is identical.
We refer to $\rvx_j$ as $\rvx$ and $\rvx_{-j}$ as $\rvz$.

Every \gls{crt} method assumes the ability to sample from $p(\rvx \mid \rvz)$ but in some of our experiments we
do not allow access to this distribution.
\Gls{diet} with \glspl{mdn} can directly model $p(\rvx \mid \rvz)$, so its approximation can be used to sample null datasets $\Dxtyz$.
However, neither the \glspl{dcrt} nor the \gls{hrt} have this facility.
For these models, we use deep generative models   to sample from $p(\rvx \mid \rvz)$ \citep{romano2020deep,sudarshan2020deep,jordon2018knockoffgan}.

\citet{romano2020deep} train a generative model $\qknockoff(\rvxt, \rvzt \mid \rvx, \rvz)$ from samples of $(\rvx, \rvz)$, which models $(\rvxt, \rvzt) \mid (\rvx, \rvz)$, where $\rvxt$ and $\rvzt$ are random variables that satisfy the following property:
\begin{align}
    [\rvxt, \rvzt, \rvx, \rvz] \disteq [\rvx, \rvzt, \rvxt, \rvz] \disteq [\rvxt, \rvz, \rvx, \rvzt] \disteq [\rvx, \rvz, \rvxt, \rvzt]. \label{eqn:swap-property} \tag{swap property}
\end{align}
The model $\qknockoff$ can then be used to generate a null dataset $\Dxtyz$.
The $i$th sample of $\rvxt$ in $\Dxtyz$ is sampled by drawing $\rvxti, \rvzt^{(i)}$ from $\qknockoff(\rvxt, \rvzt \mid \rvx = \rvxi, \rvz = \rvzi)$, then discarding $\rvzt^{(i)}$.
Due to the \ref{eqn:swap-property}, the sample $\rvxti \mid \rvzi \disteq \rvxi \mid \rvzi$, but is conditionally independent of $\rvyi \mid \rvzi$.
This makes $\rvxti$ drawn from $\qknockoff$ a valid null sample when used in each Model-X method's $p$-value computation.
The null datasets $\{ \Dxtyz^{(m)} \}_{m=1}^M$ can be drawn the same way.

It is critical to note that if type-1 error is to be controlled using the conditions laid out by \cref{prop:type-1-error}, sample splitting is required.
Since the proof of \cref{prop:type-1-error} requires that the same function $W$ be applied to the sequence
\begin{align*}
	W(\Dxyz), W(\Dxtyz^{(1)}), \dots, W(\Dxtyz^{(M)}),
\end{align*}
any estimator for $p(\rvx \mid \rvz)$ must be fit using a separate dataset.
As such, we split the dataset $\Dxyz$ into a train set $\Dxyz^{(\textrm{train})}$ and a test set $\Dxyz^{(\textrm{test})}$.
We fit models for $p(\rvx \mid \rvz)$ and the \gls{hrt} model $\qmodel$ using the training set, then compute $p$-values using the test set.

\subsection{Synthetic \gls{cvs} experiments setup}
\label{sec:synthetic-cvs-details}

In this section, we provide exact simulation details for our synthetic \gls{cvs} experiments.

The $\rvx$ data is sampled as follows: $\rvx \sim \sum_{k=1}^4 \pi_k \mathcal{N}(\mu_k \cdot \mathbf{1}, \Sigma_k)$ is a mixture of autoregressive Gaussians.
Each $\Sigma_k$ is a 100-dimensional covariance matrix whose $(i,j)$th entry is $\rho_k^{|i-j|}$.
We set $(\rho_1, \rho_2, \rho_3, \rho_4) = (0.7, 0.6, 0.5, 0.4)$, $(\pi_1, \pi_2, \pi_3, \pi_4) = (0.4, 0.3, 0.2, 0.1)$, and $(\mu_1, \mu_2, \mu_3, \mu_4) = (0, 20, 40, 60)$.

The response $\rvy \mid \rvx \sim \mathcal{N}(\langle \rvx, \beta \rangle, 1)$, where $\beta$ is a coefficient vector.
Each non-zero element of $\beta$ is drawn from $3 \cdot \text{Rademacher}(0.5)$; there are 20 non-zero elements chosen randomly in each run.
These non-zero elements represent the important variables each method aims to recover.

\subsection{Semi-synthetic genetics experiments setup}
\label{sec:semisynthetic-cvs-details}

In this section, we provide exact simulation details for our semi-synthetic genetics experiments.

To generate each dataset $\Dxyz \in \sR^{963 \times 100}$, we first sample a set of genes 100 $\{ \rvx_j \}_{j=1}^{100}$ from a set of 20K.
Let $O$ be the running set of genes, and $S$ be the full set of 20K genes.
The first gene $\rvx_1$ is sampled uniformly from $S$ and added to $O$, and removed from $S$.
For each $j > 1$, we apply the following procedure.
A gene $\rvx_k$ is drawn uniformly from $O$.
The correlation between $\rvx_j$ and each gene in $S$ is computed and the top 50 strongest correlated genes $F$ are selected.
The gene $\rvx_j \sim \mathrm{Uniform}(F)$, and is added to $O$ and removed from $S$.
This process is repeated until $S$ contains 100 genes.

To sample $\rvy \mid \rvx$, we apply the following procedure defined by \citet{liang2018bayesian}.
 The response has four main parts: two first order terms, a second order term, and a final nonlinearity term.
 
\begin{align*}
		k &\in [m / 4] \\
		\varphi^{(1)}_k, \varphi^{(2)}_k &\sim \mathcal{N}(1, 1) \\
		\varphi^{(3)}_k, \varphi^{(4)}_k, \varphi^{(5)}_k, \varphi^{(6)}_k &\sim \mathcal{N}(2, 1)
\end{align*}
\begin{align*}
	\rvy \mid \rvx &= \epsilon + \sum_{k=1}^{m / 4} 
	\varphi^{(1)}_k \rvx_{4k - 3} + 
	\varphi^{(3)}_k \rvx_{4k - 2} + 
	\varphi^{(4)}_k \rvx_{4k - 3} \rvx_{4k - 2} + 
	\varphi^{(5)}_k \tanh (\varphi^{(2)}_k \rvx_{4k - 1} + \varphi^{(6)}_k \rvx_{4k}).
\end{align*}

The variable $m$ determines the number of important features.
We set $m$ to 20 in our experiments.

\subsection{Electronic health records experiment}
\label{sec:additional-figures}

See \cref{tab:covid} for a list of features.

\begin{table}[t]
\centering
\begin{tabular}{lccccc}
\hline
Feature             & \gls{diet}      & \gls{hrt}       & $d_0$-\gls{crt}    & $d_I$-\gls{crt}    & Reference(s) \\ \hline
Age                 & $\bullet$ & $\bullet$ & $\bullet$ & $\bullet$ & (a, b, c, d, e, f, g, h) \\
Sex                 & $\bullet$ & $\bullet$ & $\bullet$ & $\bullet$ &  \\
BMI                 & $\bullet$ & $\bullet$ &           &           & (a, b) \\
Race                 &  &  &           &           &  \\
Weight              &           & $\bullet$ &           & $\bullet$ &  \\
Temperature     &  &           &           &           & \\
Heart rate          & $\bullet$ &           &           &           & (a) \\
Smoker          &  &           &           &           &  \\
Lymphocytes count          &  &           &           &           &  (g, h) \\
Lymphocytes percent          &  &           &           &           &  \\
Days since admission     & $\bullet$ &    $\bullet$       &           &      $\bullet$     &  (g) \\
Respiratory rate   &           &           &           & $\bullet$ & (h) \\
Neutrophils count &           &           &  &  & (a) \\
Neutrophils percent &           &           &  &  & \\
Eosinophils count  &           & $\bullet$ & $\bullet$ & $\bullet$ & (d, g, h) \\
Eosinophils percent & $\bullet$ &           &           & $\bullet$ & (d) \\
Blood urea nitrogen & $\bullet$ & $\bullet$ & $\bullet$ & $\bullet$ & (c, d, g) \\
Troponin     &  &           &           &           & (a, c, d, g, h) \\
Ferritin            & $\bullet$ &           &           & $\bullet$ & (b, d, g, h) \\
Platelet volume     & $\bullet$ &           &           &           & (b, f, h) \\
Platelet count     &  &           &           &           & (g, h) \\
Creatinine     &  &           &           &           & (c) \\
Lactate dehydrogenase &           &           &  &  & (a, g, h) \\
D-dimer             & $\bullet$ & $\bullet$ &           & $\bullet$ & (a, c, d, e, h) \\
C-reactive protein  & $\bullet$ &           &           & $\bullet$ & (a, b, d, g) \\
O2 Saturation       & $\bullet$ & $\bullet$ & $\bullet$ & $\bullet$ & (a, b) \\
O2 device           &           &           & $\bullet$ & $\bullet$ &  \\
High O2 support     & $\bullet$ & $\bullet$ & $\bullet$ & $\bullet$ & (a, g) \\
On room air         &           &           & $\bullet$ & $\bullet$ &  \\ 
\hline
\end{tabular}
\caption{\textbf{\Gls{diet} with \glspl{mdn} selects many medically relevant variables in the health records task, while omitting variables that provide similar but redundant information.} This table shows which variables each method selects.
We evaluate each \gls{crt} by comparing to variables found in well-cited medical articles: (a)
 \citep{petrilli2020factors}, (b) \citep{sattar2020obesity}, (c) \citep{mei2020risk}, (d) \citep{castro2020laboratory}, (e) \citep{zhang2020d}, (f) \citep{zhong2021mean}, (g) \citep{ruan2020clinical}, (h) \citep{zhou2020clinical}.
}
\label{tab:covid}
\end{table}

%% file: diet-cdf-aistats2023.bbl
\begin{thebibliography}{50}
\providecommand{\natexlab}[1]{#1}
\providecommand{\url}[1]{\texttt{#1}}
\expandafter\ifx\csname urlstyle\endcsname\relax
  \providecommand{\doi}[1]{doi: #1}\else
  \providecommand{\doi}{doi: \begingroup \urlstyle{rm}\Url}\fi

\bibitem[Bates et~al.(2020)Bates, Sesia, Sabatti, and
  Cand{\`e}s]{bates2020causal}
S.~Bates, M.~Sesia, C.~Sabatti, and E.~Cand{\`e}s.
\newblock Causal inference in genetic trio studies.
\newblock \emph{Proceedings of the National Academy of Sciences}, 117\penalty0
  (39):\penalty0 24117--24126, 2020.

\bibitem[Bellot and van~der Schaar(2019)]{bellot2019conditional}
A.~Bellot and M.~van~der Schaar.
\newblock Conditional independence testing using generative adversarial
  networks.
\newblock \emph{Advances in Neural Information Processing Systems},
  32:\penalty0 2202--2211, 2019.

\bibitem[Benjamini and Hochberg(1995)]{benjamini1995controlling}
Y.~Benjamini and Y.~Hochberg.
\newblock Controlling the false discovery rate: a practical and powerful
  approach to multiple testing.
\newblock \emph{Journal of the Royal statistical society: series B
  (Methodological)}, 57\penalty0 (1):\penalty0 289--300, 1995.

\bibitem[Benjamini and Yekutieli(2001)]{benjamini2001control}
Y.~Benjamini and D.~Yekutieli.
\newblock The control of the false discovery rate in multiple testing under
  dependency.
\newblock \emph{Annals of statistics}, pages 1165--1188, 2001.

\bibitem[Bhattacharya and Gangopadhyay(1990)]{bhattacharya1990kernel}
P.~K. Bhattacharya and A.~K. Gangopadhyay.
\newblock Kernel and nearest-neighbor estimation of a conditional quantile.
\newblock \emph{The Annals of Statistics}, pages 1400--1415, 1990.

\bibitem[Bishop(1994)]{bishop1994mixture}
C.~M. Bishop.
\newblock Mixture density networks.
\newblock 1994.

\bibitem[Cand{\`e}s et~al.(2018)Cand{\`e}s, Fan, Janson, and
  Lv]{candes2018panning}
E.~Cand{\`e}s, Y.~Fan, L.~Janson, and J.~Lv.
\newblock Panning for gold:‘model-x’knockoffs for high dimensional
  controlled variable selection.
\newblock \emph{Journal of the Royal Statistical Society: Series B (Statistical
  Methodology)}, 80\penalty0 (3):\penalty0 551--577, 2018.

\bibitem[Castro et~al.(2020)Castro, McCoy, and Perlis]{castro2020laboratory}
V.~M. Castro, T.~H. McCoy, and R.~H. Perlis.
\newblock Laboratory findings associated with severe illness and mortality
  among hospitalized individuals with coronavirus disease 2019 in eastern
  massachusetts.
\newblock \emph{JAMA network open}, 3\penalty0 (10):\penalty0
  e2023934--e2023934, 2020.

\bibitem[Cheng et~al.(1998)Cheng, Bell, and Liu]{cheng1998learning}
J.~Cheng, D.~Bell, and W.~Liu.
\newblock Learning bayesian networks from data: An efficient approach based on
  information theory.
\newblock \emph{On World Wide Web at http://www. cs. ualberta. ca/\~{}
  jcheng/bnpc. htm}, 1998.

\bibitem[Daudin(1980)]{daudin1980partial}
J.~Daudin.
\newblock Partial association measures and an application to qualitative
  regression.
\newblock \emph{Biometrika}, 67\penalty0 (3):\penalty0 581--590, 1980.

\bibitem[De~Campos and Huete(2000)]{de2000new}
L.~M. De~Campos and J.~F. Huete.
\newblock A new approach for learning belief networks using independence
  criteria.
\newblock \emph{International Journal of Approximate Reasoning}, 24\penalty0
  (1):\penalty0 11--37, 2000.

\bibitem[Doran et~al.(2014)Doran, Muandet, Zhang, and
  Sch{\"o}lkopf]{doran2014permutation}
G.~Doran, K.~Muandet, K.~Zhang, and B.~Sch{\"o}lkopf.
\newblock A permutation-based kernel conditional independence test.
\newblock In \emph{UAI}, pages 132--141. Citeseer, 2014.

\bibitem[Fukumizu et~al.(2007)Fukumizu, Gretton, Sun, and
  Sch{\"o}lkopf]{fukumizu2007kernel}
K.~Fukumizu, A.~Gretton, X.~Sun, and B.~Sch{\"o}lkopf.
\newblock Kernel measures of conditional dependence.
\newblock \emph{Advances in neural information processing systems}, 20, 2007.

\bibitem[Gretton et~al.(2012)Gretton, Borgwardt, Rasch, Sch{\"o}lkopf, and
  Smola]{gretton2012kernel}
A.~Gretton, K.~M. Borgwardt, M.~J. Rasch, B.~Sch{\"o}lkopf, and A.~Smola.
\newblock A kernel two-sample test.
\newblock \emph{The Journal of Machine Learning Research}, 13\penalty0
  (1):\penalty0 723--773, 2012.

\bibitem[Guo and Small(2016)]{guo2016control}
Z.~Guo and D.~S. Small.
\newblock Control function instrumental variable estimation of nonlinear causal
  effect models.
\newblock \emph{The Journal of Machine Learning Research}, 17\penalty0
  (1):\penalty0 3448--3482, 2016.

\bibitem[Imbens and Newey(2009)]{imbens2009identification}
G.~W. Imbens and W.~K. Newey.
\newblock Identification and estimation of triangular simultaneous equations
  models without additivity.
\newblock \emph{Econometrica}, 77\penalty0 (5):\penalty0 1481--1512, 2009.

\bibitem[Jordon et~al.(2018)Jordon, Yoon, and van~der
  Schaar]{jordon2018knockoffgan}
J.~Jordon, J.~Yoon, and M.~van~der Schaar.
\newblock Knockoffgan: Generating knockoffs for feature selection using
  generative adversarial networks.
\newblock In \emph{International Conference on Learning Representations}, 2018.

\bibitem[Katsevich and Ramdas(2020)]{katsevich2020theoretical}
E.~Katsevich and A.~Ramdas.
\newblock A theoretical treatment of conditional independence testing under
  model-x.
\newblock \emph{arXiv preprint arXiv:2005.05506}, 2020.

\bibitem[Kingma and Ba(2014)]{kingma2014adam}
D.~P. Kingma and J.~Ba.
\newblock Adam: A method for stochastic optimization.
\newblock \emph{arXiv preprint arXiv:1412.6980}, 2014.

\bibitem[Kingma et~al.(2015)Kingma, Salimans, and
  Welling]{kingma2015variational}
D.~P. Kingma, T.~Salimans, and M.~Welling.
\newblock Variational dropout and the local reparameterization trick.
\newblock \emph{arXiv preprint arXiv:1506.02557}, 2015.

\bibitem[Koller and Friedman(2009)]{koller2009probabilistic}
D.~Koller and N.~Friedman.
\newblock \emph{Probabilistic graphical models: principles and techniques}.
\newblock MIT press, 2009.

\bibitem[Lauritzen(1996)]{lauritzen1996graphical}
S.~L. Lauritzen.
\newblock \emph{Graphical models}, volume~17.
\newblock Clarendon Press, 1996.

\bibitem[Lee and Honavar(2017)]{lee2017kernel}
S.~Lee and V.~Honavar.
\newblock A kernel conditional independence test for relational data.
\newblock In \emph{33rd Conference on Uncertainty in Artificial Intelligence,
  UAI 2017}, 2017.

\bibitem[Li and Racine(2008)]{li2008nonparametric}
Q.~Li and J.~S. Racine.
\newblock Nonparametric estimation of conditional cdf and quantile functions
  with mixed categorical and continuous data.
\newblock \emph{Journal of Business \& Economic Statistics}, 26\penalty0
  (4):\penalty0 423--434, 2008.

\bibitem[Liang et~al.(2018)Liang, Li, and Zhou]{liang2018bayesian}
F.~Liang, Q.~Li, and L.~Zhou.
\newblock Bayesian neural networks for selection of drug sensitive genes.
\newblock \emph{Journal of the American Statistical Association}, 113\penalty0
  (523):\penalty0 955--972, 2018.

\bibitem[Liu et~al.(2020)Liu, Katsevich, Janson, and Ramdas]{liu2020fast}
M.~Liu, E.~Katsevich, L.~Janson, and A.~Ramdas.
\newblock Fast and powerful conditional randomization testing via distillation.
\newblock \emph{arXiv preprint arXiv:2006.03980}, 2020.

\bibitem[Mei et~al.(2020)Mei, Weinberg, Zhao, Frink, Qi, Behdad, and
  Ji]{mei2020risk}
Y.~Mei, S.~E. Weinberg, L.~Zhao, A.~Frink, C.~Qi, A.~Behdad, and P.~Ji.
\newblock Risk stratification of hospitalized covid-19 patients through
  comparative studies of laboratory results with influenza.
\newblock \emph{EClinicalMedicine}, 26:\penalty0 100475, 2020.

\bibitem[Patra et~al.(2016)Patra, Sen, and Sz{\'e}kely]{patra2016nonparametric}
R.~K. Patra, B.~Sen, and G.~J. Sz{\'e}kely.
\newblock On a nonparametric notion of residual and its applications.
\newblock \emph{Statistics \& Probability Letters}, 109:\penalty0 208--213,
  2016.

\bibitem[Pearl(2009)]{pearl2009causal}
J.~Pearl.
\newblock Causal inference in statistics: An overview.
\newblock \emph{Statistics surveys}, 3:\penalty0 96--146, 2009.

\bibitem[Petrilli et~al.(2020)Petrilli, Jones, Yang, Rajagopalan, O’Donnell,
  Chernyak, Tobin, Cerfolio, Francois, and Horwitz]{petrilli2020factors}
C.~M. Petrilli, S.~A. Jones, J.~Yang, H.~Rajagopalan, L.~O’Donnell,
  Y.~Chernyak, K.~A. Tobin, R.~J. Cerfolio, F.~Francois, and L.~I. Horwitz.
\newblock Factors associated with hospital admission and critical illness among
  5279 people with coronavirus disease 2019 in new york city: prospective
  cohort study.
\newblock \emph{Bmj}, 369, 2020.

\bibitem[Puli and Ranganath(2020)]{puli2020general}
A.~Puli and R.~Ranganath.
\newblock General control functions for causal effect estimation from ivs.
\newblock \emph{Advances in Neural Information Processing Systems}, 33, 2020.

\bibitem[Razavian et~al.(2020)Razavian, Major, Sudarshan, Burk-Rafel, Stella,
  Randhawa, Bilaloglu, Chen, Nguy, Wang, et~al.]{razavian2020validated}
N.~Razavian, V.~J. Major, M.~Sudarshan, J.~Burk-Rafel, P.~Stella, H.~Randhawa,
  S.~Bilaloglu, J.~Chen, V.~Nguy, W.~Wang, et~al.
\newblock A validated, real-time prediction model for favorable outcomes in
  hospitalized covid-19 patients.
\newblock \emph{NPJ digital medicine}, 3\penalty0 (1):\penalty0 1--13, 2020.

\bibitem[Romano et~al.(2020)Romano, Sesia, and Cand{\`e}s]{romano2020deep}
Y.~Romano, M.~Sesia, and E.~Cand{\`e}s.
\newblock Deep knockoffs.
\newblock \emph{Journal of the American Statistical Association}, 115\penalty0
  (532):\penalty0 1861--1872, 2020.

\bibitem[Ruan et~al.(2020)Ruan, Yang, Wang, Jiang, and Song]{ruan2020clinical}
Q.~Ruan, K.~Yang, W.~Wang, L.~Jiang, and J.~Song.
\newblock Clinical predictors of mortality due to covid-19 based on an analysis
  of data of 150 patients from wuhan, china.
\newblock \emph{Intensive care medicine}, 46\penalty0 (5):\penalty0 846--848,
  2020.

\bibitem[Runge(2018)]{runge2018conditional}
J.~Runge.
\newblock Conditional independence testing based on a nearest-neighbor
  estimator of conditional mutual information.
\newblock In \emph{International Conference on Artificial Intelligence and
  Statistics}, pages 938--947. PMLR, 2018.

\bibitem[Sattar et~al.(2020)Sattar, McInnes, and McMurray]{sattar2020obesity}
N.~Sattar, I.~B. McInnes, and J.~J. McMurray.
\newblock Obesity is a risk factor for severe covid-19 infection: multiple
  potential mechanisms.
\newblock \emph{Circulation}, 142\penalty0 (1):\penalty0 4--6, 2020.

\bibitem[Sen et~al.(2017)Sen, Suresh, Shanmugam, Dimakis, and
  Shakkottai]{sen2017model}
R.~Sen, A.~T. Suresh, K.~Shanmugam, A.~G. Dimakis, and S.~Shakkottai.
\newblock Model-powered conditional independence test.
\newblock \emph{Advances in neural information processing systems}, 30, 2017.

\bibitem[Sesia et~al.(2019)Sesia, Sabatti, and Cand{\`e}s]{sesia2019gene}
M.~Sesia, C.~Sabatti, and E.~J. Cand{\`e}s.
\newblock Gene hunting with hidden {M}arkov model knockoffs.
\newblock \emph{Biometrika}, 106\penalty0 (1):\penalty0 1--18, 2019.

\bibitem[Spirtes et~al.(2000)Spirtes, Glymour, Scheines, and
  Heckerman]{spirtes2000causation}
P.~Spirtes, C.~N. Glymour, R.~Scheines, and D.~Heckerman.
\newblock \emph{Causation, prediction, and search}.
\newblock MIT press, 2000.

\bibitem[Sudarshan et~al.(2020)Sudarshan, Tansey, and
  Ranganath]{sudarshan2020deep}
M.~Sudarshan, W.~Tansey, and R.~Ranganath.
\newblock Deep direct likelihood knockoffs.
\newblock \emph{Advances in Neural Information Processing Systems}, 33, 2020.

\bibitem[Sudarshan et~al.(2021)Sudarshan, Puli, Subramanian, Sankararaman, and
  Ranganath]{sudarshan2021contra}
M.~Sudarshan, A.~Puli, L.~Subramanian, S.~Sankararaman, and R.~Ranganath.
\newblock Contra: Contrarian statistics for controlled variable selection.
\newblock In \emph{International Conference on Artificial Intelligence and
  Statistics}, pages 1900--1908. PMLR, 2021.

\bibitem[Tansey et~al.(2022)Tansey, Veitch, Zhang, Rabadan, and
  Blei]{tansey2018holdout}
W.~Tansey, V.~Veitch, H.~Zhang, R.~Rabadan, and D.~M. Blei.
\newblock The holdout randomization test for feature selection in black box
  models.
\newblock \emph{Journal of Computational and Graphical Statistics}, 31\penalty0
  (1):\penalty0 151--162, 2022.

\bibitem[Trivedi and Zimmer(2007)]{trivedi2007copula}
P.~K. Trivedi and D.~M. Zimmer.
\newblock \emph{Copula modeling: an introduction for practitioners}.
\newblock Now Publishers Inc, 2007.

\bibitem[Vinh et~al.(2009)Vinh, Epps, and Bailey]{vinh2009information}
N.~X. Vinh, J.~Epps, and J.~Bailey.
\newblock Information theoretic measures for clusterings comparison: is a
  correction for chance necessary?
\newblock In \emph{Proceedings of the 26th annual international conference on
  machine learning}, pages 1073--1080, 2009.

\bibitem[Yang et~al.(2012)Yang, Soares, Greninger, Edelman, Lightfoot, Forbes,
  Bindal, Beare, Smith, Thompson, et~al.]{yang2012genomics}
W.~Yang, J.~Soares, P.~Greninger, E.~J. Edelman, H.~Lightfoot, S.~Forbes,
  N.~Bindal, D.~Beare, J.~A. Smith, I.~R. Thompson, et~al.
\newblock Genomics of drug sensitivity in cancer (gdsc): a resource for
  therapeutic biomarker discovery in cancer cells.
\newblock \emph{Nucleic acids research}, 41\penalty0 (D1):\penalty0 D955--D961,
  2012.

\bibitem[Zhang et~al.(2012)Zhang, Peters, Janzing, and
  Sch{\"o}lkopf]{zhang2012kernel}
K.~Zhang, J.~Peters, D.~Janzing, and B.~Sch{\"o}lkopf.
\newblock Kernel-based conditional independence test and application in causal
  discovery.
\newblock \emph{arXiv preprint arXiv:1202.3775}, 2012.

\bibitem[Zhang et~al.(2020)Zhang, Yan, Fan, Liu, Liu, Liu, and
  Zhang]{zhang2020d}
L.~Zhang, X.~Yan, Q.~Fan, H.~Liu, X.~Liu, Z.~Liu, and Z.~Zhang.
\newblock D-dimer levels on admission to predict in-hospital mortality in
  patients with covid-19.
\newblock \emph{Journal of Thrombosis and Haemostasis}, 18\penalty0
  (6):\penalty0 1324--1329, 2020.

\bibitem[Zhong and Peng(2021)]{zhong2021mean}
Q.~Zhong and J.~Peng.
\newblock Mean platelet volume/platelet count ratio predicts severe pneumonia
  of covid-19.
\newblock \emph{Journal of clinical laboratory analysis}, 35\penalty0
  (1):\penalty0 e23607, 2021.

\bibitem[Zhou et~al.(2020)Zhou, Yu, Du, Fan, Liu, Liu, Xiang, Wang, Song, Gu,
  et~al.]{zhou2020clinical}
F.~Zhou, T.~Yu, R.~Du, G.~Fan, Y.~Liu, Z.~Liu, J.~Xiang, Y.~Wang, B.~Song,
  X.~Gu, et~al.
\newblock Clinical course and risk factors for mortality of adult inpatients
  with covid-19 in wuhan, china: a retrospective cohort study.
\newblock \emph{The lancet}, 395\penalty0 (10229):\penalty0 1054--1062, 2020.

\bibitem[Zhu et~al.(2018)Zhu, Zheng, Zhang, Wu, Trzaskowski, Maier, Robinson,
  McGrath, Visscher, Wray, et~al.]{zhu2018causal}
Z.~Zhu, Z.~Zheng, F.~Zhang, Y.~Wu, M.~Trzaskowski, R.~Maier, M.~R. Robinson,
  J.~J. McGrath, P.~M. Visscher, N.~R. Wray, et~al.
\newblock Causal associations between risk factors and common diseases inferred
  from gwas summary data.
\newblock \emph{Nature communications}, 9\penalty0 (1):\penalty0 1--12, 2018.

\end{thebibliography}
